\documentclass[sigconf]{acmart}
\AtBeginDocument{%
  \providecommand\BibTeX{{%
    \normalfont B\kern-0.5em{\scshape i\kern-0.25em b}\kern-0.8em\TeX}}}

\setcopyright{acmcopyright}
\copyrightyear{2023}
\acmYear{2023}
\acmDOI{XXXXXXX.XXXXXXX}

\acmConference[CHI '24]{Conference on Human Factors in Computing Systems}{May 11--16, 2024}{Honolulu, HI, USA}
\acmBooktitle{CHI'24: Proceedings of the 2024 CHI Conference on Human Factors in Computing Systems, May 11--16, 2024, Honolulu, HI, USA} 
\acmPrice{15.00}
\acmISBN{978-1-4503-XXXX-X/18/06}

\usepackage{graphicx} 
\usepackage{environ}
\usepackage{subcaption}
\usepackage{multirow}
\usepackage{tabularx}
\usepackage{makecell}
\usepackage{tikz}
\usetikzlibrary{shapes.geometric, arrows}
\newcolumntype{?}{!{\vrule width 2pt}}
\NewEnviron{myequation}{%
\begin{equation*}
\scalebox{1.2}{$\BODY$}
\end{equation*}
}

\begin{document}

\title{HILL: A Hallucination Identifier for Large Language Models}

\author{Florian Leiser}
\orcid{0000-0001-5347-0493}
\affiliation{%
  \institution{Institute of Applied Informatics and Formal Description Methods,\\ Karlsruhe Institute of Technology}
  \city{Karlsruhe}
  \country{Germany}
  \postcode{76131}
}
\email{florian.leiser@kit.edu}

\author{Sven Eckhardt}
\orcid{0000-0002-4713-8408}
\affiliation{%
  \institution{Department of Informatics,\\ University of Zurich}
  \city{Zurich}
  \country{Switzerland}
  \postcode{8050}
}
\email{eckhardt@ifi.uzh.ch}

\author{Valentin Leuthe}
\orcid{0009-0002-8938-302X}
\affiliation{%
  \institution{Institute of Applied Informatics and Formal Description Methods,\\ Karlsruhe Institute of Technology}
  \city{Karlsruhe}
  \country{Germany}
  \postcode{76131}
}
\email{valentin.leuthe@student.kit.edu}

\author{Merlin Knaeble}
\orcid{0000-0002-5108-4609}
\affiliation{%
  \institution{Human-Centered Systems Lab,\\ Karlsruhe Institute of Technology}
  \city{Karlsruhe}
  \country{Germany}
  \postcode{76131}
}
\email{merlin.knaeble@gmail.com}

\author{Alexander Maedche}
\orcid{0000-0001-6546-4816}
\affiliation{%
  \institution{Human-Centered Systems Lab,\\ Karlsruhe Institute of Technology}
  \city{Karlsruhe}
  \country{Germany}
  \postcode{76131}
}
\email{alexander.maedche@kit.edu}

\author{Gerhard Schwabe}
\orcid{0000-0002-0453-9762}
\affiliation{%
  \institution{Department of Informatics,\\ University of Zurich}
  \city{Zurich}
  \country{Switzerland}
  \postcode{8050}
}
\email{schwabe@ifi.uzh.ch}

\author{Ali Sunyaev}
\orcid{0000-0002-4353-8519}
\affiliation{%
  \institution{Institute of Applied Informatics and Formal Description Methods,\\ Karlsruhe Institute of Technology}
  \city{Karlsruhe}
  \country{Germany}
  \postcode{76131}
}
\email{sunyaev@kit.edu}

\renewcommand{\shortauthors}{Leiser, F.; Eckhardt, S.; Leuthe, V.; Knaeble, M.; Maedche, A.; Schwabe, G.; Sunyaev, A.}
\renewcommand{\shorttitle}{Hallucination Identifier for Large Language Models}

\begin{abstract}
    Large language models (LLMs) are prone to hallucinations, i.e., nonsensical, unfaithful, and undesirable text. Users tend to overrely on LLMs and corresponding hallucinations which can lead to misinterpretations and errors. To tackle the problem of overreliance, we propose HILL, the \textit{"Hallucination Identifier for Large Language Models"}. First, we identified design features for HILL with a Wizard of Oz approach with nine participants. Subsequently, we implemented HILL based on the identified design features and evaluated HILL's interface design by surveying 17 participants. Further, we investigated HILL's functionality to identify hallucinations based on an existing question-answering dataset and five user interviews. We find that HILL can correctly identify and highlight hallucinations in LLM responses which enables users to handle LLM responses with more caution. With that, we propose an easy-to-implement adaptation to existing LLMs and demonstrate the relevance of user-centered designs of AI artifacts.
\end{abstract}





\begin{teaserfigure}
  \centering
  \includegraphics[width=0.73\textwidth]{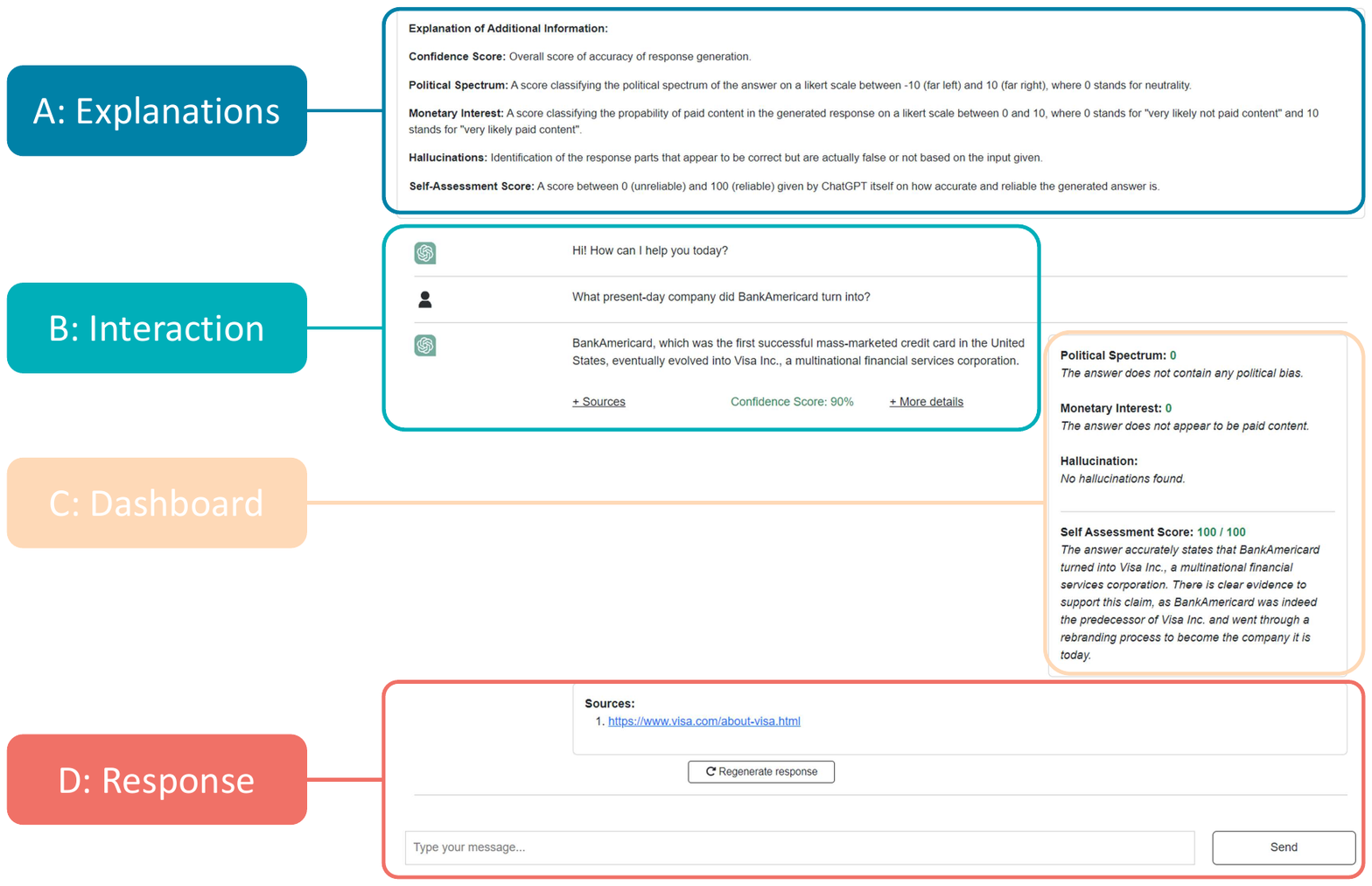}
  \caption{Screenshot of the developed HILL artifact, highlighting hallucinations to users, enabling them to assess the factual correctness of an LLM response.}
  \Description{This figure shows the interface of our developed HILL artifact. Within the artifact, four areas are marked with differently colored boxes. The top-most box is blue and has in white written "A: Explanations" in it. The box contains the explanations of the features Confidence Score, Political Spectrum, Monetary Interest, Hallucinations, and Self-Assessment Score. Below that is box "B: Interaction" which shows a user interaction with ChatGPT with the icons of OpenAI. Below this conversation is a confidence score presented in green and two buttons to open Sources or more details. The "more details" button opens the third box on the right sight labeled "C: Dashboard" in a light yellow color. In there, the features Political Spectrum, Monetary Interest, Hallucination, and Self-Assessment Score are explained. The final box "D: Response" highlights the bottom of the figure in light red where the user can type in a response, regenerate the response, or click on the sources available in the source dashboard.}
  \label{fig:frontend_prototype}
\end{teaserfigure}

\received{14 September 2023}
\received[revised]{12 December 2023}
\received[accepted]{18 January 2024}

\maketitle

\section{Introduction}

Large language models (LLMs) like OpenAI's ChatGPT \cite{openai2023} or Google's LaMDA \cite{lamda2021} have gained immense interest over the last year. With their human-like response generation, they enable everyday users to interact with generative artificial intelligence (GenAI) intuitively. However, the more users interact with LLMs, the more errors and sometimes ridiculous responses occur \cite{borji2023categorical}. Examples include contradictory statements within one reply of ChatGPT when assessing whether 5 is a prime number, wrongly explaining programmer humor, or generating factually wrong responses (e.g., assessing that 1000 is larger than 1062) \cite{borji2023categorical}. These errors stem from the probability-based nature of LLMs \cite{carlini2021extracting} which may contain human biases in the underlying data \cite{Eckhardt2023} and introduce major challenges for the everyday user interacting with LLMs. These erroneous outputs are generally referred to as hallucinations \cite{guerreiro2023hallucinations}.

In summary, hallucinations are defined as text that is nonsensical, unfaithful, and undesirable \cite{ji2023survey}. Hallucinations are not limited to the human-like response generation of LLMs. For example, hallucinations have also been talked about in abstractive text summarization \cite[e.g.,][]{maynez2020faithfulness}. These hallucinations are critical when users rely on the response of artificial intelligence (AI) models. With the increasing use of AI responses, these responses might even influence users' political beliefs \cite{kahne2018political}. Recent human-AI-collaboration efforts include domain knowledge to improve the accuracy and explainability of model responses \cite{vonrueden2023informed}. In the context of LLMs, this knowledge is included in the model structure \cite{NEURIPS2022_Ouyang}, as self-assessment of the model response \cite{manakul2023selfcheckgpt}, or in prompt engineering \cite{white2023prompt, zamfirescu2023johnny}. This knowledge-informed machine learning provides essential technical efforts to improve LLM responses, but it is unforeseeable when LLMs will be error-free \cite{narayan_elon_2023}.

Therefore, users are at risk of frequently relying on incorrect information provided by LLMs. This problem is called overreliance \cite{lai2021towards} and solutions should be incorporated during the design and development of these systems \cite{passi2022overreliance}. Since LLMs are established in the general public and handle frequently changing domains by everyday users \cite{yang_2023_LLMsinPractice}, it is of great importance to reduce the overreliance on LLM responses.

Previous studies investigated how hallucinations can be avoided to reduce overreliance in LLMs \cite{leiser_chatgpt_2023, manakul2023selfcheckgpt, peng2023check}. These studies either focus on technical approaches without incorporating user feedback \cite{manakul2023selfcheckgpt} or purely develop design recommendations for LLMs based on features urged by users \cite{leiser_chatgpt_2023}. Combining both lenses and developing user-centered artifacts is of utmost importance to involve users in the design process and incorporate their desires.

To tackle this issue, we propose a novel artifact called \textit{"Hallucination Identifier in Large Language Models"} (HILL). The user-centered design of HILL, first, builds on three prototypes developed in a Wizard of Oz study (WOz) using Figma implementing existing user-desired features \cite{leiser_chatgpt_2023}. The features in the prototypes are evaluated with think-aloud sessions and semi-structured interviews analyzed with thematic analysis \cite{braun_using_2006} and best-worst scaling \cite{Finn1992}. Building on this feature prioritization, we develop the HILL artifact building on the existing ChatGPT application programming interfaces (APIs). The artifact is in turn evaluated by surveying 17 participants, confronting HILL with 128 questions from the second version of the \textit{Stanford Question Answering Dataset} (SQuAD 2.0) and five additional interviews to investigate user reliance. We conclude this paper by highlighting the contributions of HILL as well as outlining potential avenues for future improvement.

\section{Related Work}

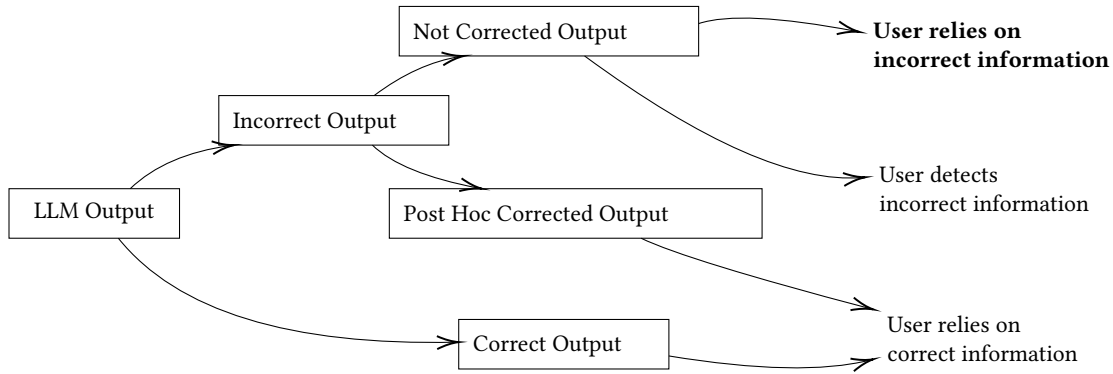
\begin{figure*}[]
    \centering
    \begin{tikzpicture}[x=0.75pt,y=0.75pt,yscale=-1,xscale=1]

\draw    (5.19,122) -- (91.19,122) -- (91.19,147) -- (5.19,147) -- cycle  ;
\draw (48.19,134.5) node   [align=left] {LLM Output};
\draw    (111,75) -- (227,75) -- (227,100) -- (111,100) -- cycle  ;
\draw (114,79) node [anchor=north west] [align=left] {Incorrect Output};
\draw    (232,188) -- (338,188) -- (338,213) -- (232,213) -- cycle  ;
\draw (235,192) node [anchor=north west] [align=left] {Correct Output};
\draw    (202,30) -- (353,30) -- (353,55) -- (202,55) -- cycle  ;
\draw (205,34) node [anchor=north west] [align=left] {Not Corrected Output};
\draw    (197,122) -- (385,122) -- (385,147) -- (197,147) -- cycle  ;
\draw (200,126) node [anchor=north west] [align=left] {Post Hoc Corrected Output};
\draw (437,33) node [anchor=north west] [align=left] {\textbf{User relies on}\\\textbf{incorrect information}};
\draw (440,106) node [anchor=north west] [align=left] {User detects\\incorrect information};
\draw (444,182) node [anchor=north west] [align=left] {User relies on \\correct information};
\draw    (66.33,122) .. controls (78.4,110.43) and (95.28,103.16) .. (116.97,100.22) ;
\draw [shift={(118.65,100)}, rotate = 172.93] [color={rgb, 255:red, 0; green, 0; blue, 0 }  ][line width=0.75]    (10.93,-3.29) .. controls (6.95,-1.4) and (3.31,-0.3) .. (0,0) .. controls (3.31,0.3) and (6.95,1.4) .. (10.93,3.29)   ;
\draw    (60.33,147) .. controls (89.96,183.82) and (146.76,201.3) .. (230.73,199.46) ;
\draw [shift={(232,199.43)}, rotate = 178.61] [color={rgb, 255:red, 0; green, 0; blue, 0 }  ][line width=0.75]    (10.93,-3.29) .. controls (6.95,-1.4) and (3.31,-0.3) .. (0,0) .. controls (3.31,0.3) and (6.95,1.4) .. (10.93,3.29)   ;
\draw    (189.35,75) .. controls (203.7,64.18) and (218.17,57.61) .. (232.79,55.27) ;
\draw [shift={(234.61,55)}, rotate = 172.26] [color={rgb, 255:red, 0; green, 0; blue, 0 }  ][line width=0.75]    (10.93,-3.29) .. controls (6.95,-1.4) and (3.31,-0.3) .. (0,0) .. controls (3.31,0.3) and (6.95,1.4) .. (10.93,3.29)   ;
\draw    (188.47,100) .. controls (198.36,109.55) and (216.2,116.78) .. (242,121.7) ;
\draw [shift={(243.59,122)}, rotate = 190.48] [color={rgb, 255:red, 0; green, 0; blue, 0 }  ][line width=0.75]    (10.93,-3.29) .. controls (6.95,-1.4) and (3.31,-0.3) .. (0,0) .. controls (3.31,0.3) and (6.95,1.4) .. (10.93,3.29)   ;
\draw    (353,38.66) .. controls (367.45,33.95) and (393.86,35.24) .. (432.24,42.53) ;
\draw [shift={(434,42.87)}, rotate = 190.92] [color={rgb, 255:red, 0; green, 0; blue, 0 }  ][line width=0.75]    (10.93,-3.29) .. controls (6.95,-1.4) and (3.31,-0.3) .. (0,0) .. controls (3.31,0.3) and (6.95,1.4) .. (10.93,3.29)   ;
\draw    (295.47,55) .. controls (354.91,98.67) and (401.58,119.12) .. (435.46,116.36) ;
\draw [shift={(437,116.22)}, rotate = 174.08] [color={rgb, 255:red, 0; green, 0; blue, 0 }  ][line width=0.75]    (10.93,-3.29) .. controls (6.95,-1.4) and (3.31,-0.3) .. (0,0) .. controls (3.31,0.3) and (6.95,1.4) .. (10.93,3.29)   ;
\draw    (323.99,147) .. controls (339.84,154.59) and (378.22,166.4) .. (439.15,182.43) ;
\draw [shift={(441,182.92)}, rotate = 194.71] [color={rgb, 255:red, 0; green, 0; blue, 0 }  ][line width=0.75]    (10.93,-3.29) .. controls (6.95,-1.4) and (3.31,-0.3) .. (0,0) .. controls (3.31,0.3) and (6.95,1.4) .. (10.93,3.29)   ;
\draw    (338,206.42) .. controls (382.2,213.63) and (415.95,214.34) .. (439.23,208.56) ;
\draw [shift={(441,208.1)}, rotate = 164.95] [color={rgb, 255:red, 0; green, 0; blue, 0 }  ][line width=0.75]    (10.93,-3.29) .. controls (6.95,-1.4) and (3.31,-0.3) .. (0,0) .. controls (3.31,0.3) and (6.95,1.4) .. (10.93,3.29)   ;

\end{tikzpicture}
    \caption{The positioning and delimitation of this study. While current approaches focus on correcting LLM hallucinations, we aim at empowering the users to not rely on hallucinations (bold). These approaches are not competing against each other but rather complementing and both are important to achieve the unified goal of users not blindly following incorrect LLM output.}
    \label{fig:positioning}
    \Description{This figure shows different processing possibilities of LLM output for users from left to right. Based on LLM outputs, two possibilities of the factual accuracy of the response exist, incorrect or correct output. For correct output, another arrow is shown indicating the user relies on correct information. For incorrect output, again, two possibilities exist, either the output is corrected or it is not. For post-hoc corrected output, the user identifies incorrect information. If the output is not correct, either the user has to detect the incorrect information themselves or they rely on incorrect information which is called overreliance.}
\end{figure*}

Probability-based AI systems are particularly opaque and prone to inexplicable errors \cite{holzinger2018machine}. Therefore, hallucinations have been a long observed issue in AI models, for example in abstractive text summarization \cite{maynez2020faithfulness}. One strategy to reduce hallucinations is to limit the summary to general phrases, which would greatly reduce the informativeness \cite{zhao2020reducing}. Another case of hallucinations occur in machine translation \cite{raunak2021curious}. Finally, hallucinations are not only known in natural language processing (NLP) but also in computer vision \cite{zhou2023analyzing}. However, in this study, we use hallucinations in the field of NLP. Hallucinations are defined as text that is nonsensical, unfaithful, and undesirable \cite{ji2023survey}. Recent research further classifies hallucinations as input-conflicting hallucination, context-conflicting hallucination, and fact-conflicting hallucination \cite{zhang2023siren}. In this study, we summarize all three classes as hallucinations as done by most current research.

With the newest advances in LLMs, hallucinations are more timely than ever. For example, GPT-4 still suffers from hallucinations \cite{openai2023gpt4}. Current research focuses on identifying and raising awareness of the errors made by LLMs \cite{borji2023categorical}. This discussion also spills over to grey literature, such as news articles and blog posts \cite[e.g.,][]{brereton_bing_2023,bell_fake_2023,vincent_google_2023,coulter_alphabet_2023}. Therefore, it is of utmost importance to ensure that users recognize potential errors and do not blindly follow the output of LLMs.

The problem of blindly following incorrect information is known as overreliance in AI systems \cite[e.g.,][]{parasuraman1997humans, lai2021towards}. This overreliance must be considered when developing and designing these systems \cite{passi2022overreliance}. Reducing overreliance can come in many ways like auditing, benchmark tests, or data quality reviews \cite{shneiderman2020human}. The goal for designers and developers of AI-based systems should be to achieve appropriate reliance of the users (i.e., a user follows correct and does not follow incorrect output) \cite{lee2004trust,schemmer2023appropriate}. Overall, previous research on LLMs focused on reducing model-made errors. However, it is not foreseeable when these models will be error-free despite the relevance of non-maleficence to achieve trustworthy AI \cite{Thiebes2020}. Instead of reducing errors, we aim to mitigate the effect of errors on the user side by empowering users to detect LLM hallucinations. Some recent approaches exist that aim at post-hoc improving the LLM output by fact-checking. One example is a "Truth-O-Meter", that fact-checks LLM output using web mining \cite{galitsky2023truth}. Another example is a so-called "LLM-Augmenter" \cite{peng2023check}, augmenting the LLM output using external knowledge. Further, a "Retrofit Attribution using Research and Revision" method can improve LLM output \cite{gao2023rarr}. 

The unified goal of all current approaches is to improve the model performance after the response generation. While these technical pursuits are worthwhile, we see fit for another angle addressing this problem. Instead of improving LLM responses, we aim to empower users to detect hallucinations themselves by using design features for LLM interfaces. Recent literature introduced some initial design aspects that may help users detect hallucinations \cite{leiser_chatgpt_2023}. However, these studies remain on the conceptual level without implementing an actual user-centered design. In this study, we implement some of the features in an artifact that is evaluated with everyday users. This difference in approach is illustrated in \autoref{fig:positioning}. While current approaches aim at correcting hallucinations, we aim at empowering the user to detect hallucinations by themselves. It is important to note that these approaches are not competing against each other, but rather complementing each other. 

\section{Study 1: Wizard of Oz}
\label{sec:woz}

We, first, conducted a WOz to instantiate and prioritize the features for an LLM interface enabling users to identify potential hallucinations. To allow comparison between the prototypes, we included one feature per category in all prototypes. We, therefore, generated three prototypes containing five to seven features each. These prototypes were then pre-tested with two user experience (UX) experts and one everyday user. The UX experts both work in an HCI-heavy context, especially with conversational agents. We, finally, conducted nine think-aloud sessions where the participants engaged with all three prototypes and followed the think-aloud sessions with post-hoc questionnaires and interviews. The combined sessions were analyzed using thematic analysis \cite{braun_using_2006}. Based on the recommendations of the UX experts, we included questionnaires regarding the system usability scale (SUS) \cite{bangor2008empirical} and conducted best-worst scaling on the features per category \cite{Finn1992}. We will present each step now in more detail.

\subsection{Prototype Development}

\begin{table*}[]
    \centering
    \begin{tabularx}{0.95\textwidth}{l*{3}{X}}
        \toprule
        \textbf{Category} & \textbf{Group 1} & \textbf{Group 2} & \textbf{Group 3}  \\\midrule
        \textbf{Confidence Score} &  \makecell[tl]{Colored Ordinal CS} & Ordinal CS & Metric CS \\
        \textbf{Sources} & \makecell[tl]{Source Links (Drill-Down),\\ Source Quality} & Direct Quotes & \makecell[tl]{Source Links (Pop-up), \\ Source Quantity}\\
        \textbf{Disclosure} & Monetary Interest & Disclaimer & \makecell[tl]{Political Spectrum,\\ Ethical \& Legal Consid.} \\
        \textbf{Visual Aid} & Response Type & - & Color\\
        \textbf{In Text} & Explanations & Phrasing & - \\
        \textbf{Customization} & Replies & Confidence Threshold & Drill-down Dashboard\\\bottomrule
    \end{tabularx}
    \caption{Distribution of 17 features of \cite{leiser_chatgpt_2023} into three groups.}
    \label{tab:feature_grouping}
\end{table*}

The design features for the prototypes are based on the introduced features in \cite{leiser_chatgpt_2023}. We split the features into three prototypes with one feature per category per group which led to a distinct set of design features, as indicated in \autoref{tab:feature_grouping}. Since not all categories contain three features, some groups have none or more than one design feature per category. Based on the three groups of features, we developed three prototypical interfaces with Figma\footnote{\url{https://www.figma.com/}, Version: 116.12.2} which enabled us to evaluate the features within each category. We built upon the widely known interface of ChatGPT and imitated a question and a corresponding LLM-generated response. An exemplary screenshot of one prototype is presented in \autoref{fig:mockup_1}, and the remaining prototypes are presented in supplement material.

\begin{figure*}[]
    \centering
    \includegraphics[width = 0.75\textwidth]{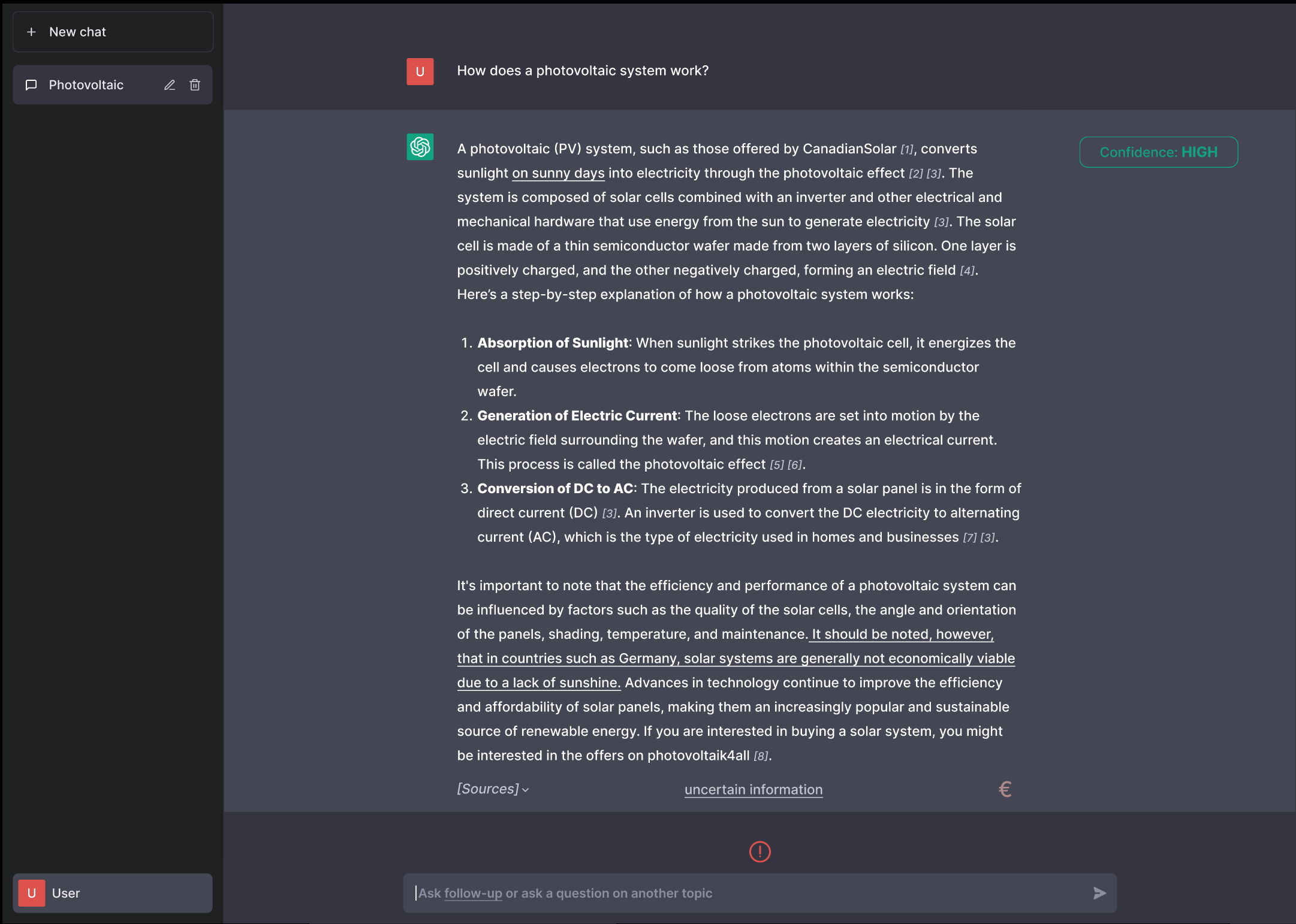}
    \caption{Screenshot of the prototypical interface including the features of Group 1.}
    \label{fig:mockup_1}
    \Description{Presentation of a ChatGPT interface in dark mode. The left side shows a small menu where new chats can be opened and the current topic "Photovoltaic" is highlighted. In the main frame, the user poses a question about the functionality of photovoltaic systems with a response of ChatGPT. In this response some information are underlined, indicating uncertain information, and the most relevant information are presented in bold. In the top right corner is a green box indicating a high confidence in response generation. Below the response are two buttons to assess the sources shown in this reply as well as a currency symbol indicating paid content in the response.}
\end{figure*}

\paragraph{Interface Descriptions.}
Each interface shows a user interaction with ChatGPT. In the interaction, a pre-determined question is posed to the LLM and the response is augmented with a feature group specified in \autoref{tab:feature_grouping}. In the first interface, the user poses a question regarding a photovoltaic system where seven features are included. We represent a colored ordinal confidence score (CS) with a green "HIGH" label in the top right corner of the response. We include source links in the response, presented in IEEE citation style. On the lower left of the response is a drill-down menu for the sources where the links to the citations are provided as well as an indicator of the source quality on a continuous scale between "low" and "high". In this sample, the source quality is specified as slightly above "medium". A currency symbol is shown on the right side below the response and highlighted in light red. When hovering over this symbol, two text paragraphs are highlighted in the same shade indicating potential monetary interest in the response generation. Other paragraphs in the response are underlined indicating uncertain information and therefore, a different response type, in the text. This explanation is given centrally below the response. An explanation of the response generation is displayed behind a red exclamation mark above the text input field. It refers to the validity and selection of sources as well as the limited timeliness and functionality of the LLM and encourages users to question the response. Finally, the users can interact with the response and pose a pre-determined follow-up question. The question asks for additional information and a second response elaborates on the previous "imprecise" one.

In the second interface, the user requests a short summary of the Cold War. When starting this conversation for the first time, a disclaimer is presented as a pop-up window explaining LLMs are "not always right" and the responses are based on statistical approximations and predictions. Another feature included in this interface is a response confidence threshold slider where the user could specify the minimal confidence in the model's response generation on a scale between 5\% and 95\%. The user can interact with the slider even after the initial response is generated which changes the response output. This interface contains another ordinal CS in the top right corner without colored representation. The score changes between "HIGH", "MEDIUM", or "LOW", depending on the minimum confidence threshold. Also depending on the threshold is the generated response that includes more or less passages of direct quotes. The corresponding sources are again presented in IEEE citation style and can be accessed by clicking on a drill-down menu in the bottom left corner of the response. Within the response with lower minimum confidence, we changed the phrasing of the response to emphasize sentences where the model was uncertain, for example, by including attenuation like "I'm not entirely sure, but ..." or "To my knowledge, ...".

Our third interface mimics an interaction with ChatGPT on a medical question regarding a cough. In this interface, we include a metric CS assessing the confidence of the response. We provide sources in the response in the same way by including them as citations in IEEE style. When investigating the sources overview, a pop-up window (instead of the previous drill-down menus) opens stating the overall number of sources used in the reply (four in this interface) as well as links to the sources. We also include the political orientation of the response by using a scale between "far left" and "far right". The interface additionally showed ethical and legal considerations where we emphasize the general functionality of LLMs and that their built on general knowledge. Additional pieces of information are provided about potential biases in response generation and the lack of human empathy in statistical-based models. For legal considerations, we specify information about personal rights, data protection, and copyright infringements. Since the interaction concerns a medical question, we also include information about the medical expertise of ChatGPT and that the user should consult a medical doctor "for an accurate diagnosis and [an] appropriate treatment plan". All this information is hidden in a drill-down dashboard. Furthermore, similar to the first interface, some passages with uncertain information are underlined, but this time in red. When clicking on the uncertain information button in the lower right corner, the passages are highlighted in the same color.

\paragraph{Pre-tests.} Following the development of the three prototypical interfaces, we conducted three pre-tests with two UX experts (both male, average age 30.0) and one everyday user (female, age 23). The UX experts frequently work on designing intelligent user interfaces, one of them primarily for data representation and the other for chatbots and LLMs specifically. The pre-tests took about 30-45 minutes each where they assessed all three prototypes. During the pre-tests, we focused on the depiction of the features and their intended functionality rather than the benefits of said features. Overall, the resonance in these pre-tests was very positive and the UX experts were keen on the results as well. They had smaller suggestions to further improve the depiction of the political orientation feature. One pre-tester suggested using a tachometer instead of a scale to ease the understanding of the political orientation. Therefore, we updated the representation to a tachometer ranging from "far left" to "far right" with a pointer indicating a rather central opinion. Based on other feedback, we extended the functionality of the response type to highlight all underlined text when the user clicks on it. We also changed the layout of the dashboard in the third prototype because one of the pre-testers suggested a tab system including political, ethical, and legal considerations. Two of the three pre-testers had some misconceptions regarding the \textit{confidence threshold} in interface two. They wondered which benefits a response with low generation probability could have. To avoid future confusion, we included a descriptive text in a question mark box just above the confidence threshold slider explaining that low generation probability might be desirable in rather creative LLM tasks.

\subsection{Qualitative Prototype Evaluation}

We evaluated the prototypes based on a WOz, where human-computer interaction is simulated \cite{fraser1991simulating}. The system is controlled by a human wizard, which is unknown to the user who believes the system is real \cite{riek2012wizard}. In our study, the prototypes operate independently, therefore, the wizard only intervenes when the prototypes fail or their usage is unclear to the participant \cite{dow2005wizard}. We wrote a short introductory text including general knowledge about LLMs and a high-level description of the participants' task before the participants were free to interact with the system \cite{riek2012wizard}. The participants were shown the three prototypes in random order. We concluded the nine WOz sessions with semi-structured interviews to evaluate the prioritization, usefulness, and usability of the features \cite{fraser1991simulating}. The interview guideline covered the opening, introduction, key questions, and closing of the interview \cite{myers2007qualitative}. We asked key questions and suggestions for improvement regarding the three most helpful features (top features), features without perceived value, or difficulties in understanding the features. An overview of the participants' top features as well as some exemplary quotes are shown in \autoref{tab:woz_quotes}.

\begin{table*}[tbp!]
    \centering
    \begin{tabularx}{\textwidth}{clXc}
        \toprule
        & \textbf{Feature} & \textbf{Exemplary Quote} & \textbf{Top} \\\midrule
        \multirow{7}{*}{\rotatebox{90}{\textbf{Group 1}}} & Colored Ordinal CS & \textit{"traffic light system [... is] immediately understandable"} [p02] & 3 \\
        & Source Links (Drill-Down) &  \textit{"see at a quick glance [...] where the information comes from"} [p04] & 6 \\
        & Source Quality &  \textit{"an easy indicator of how good the sources are"} [p01] & 3 \\
        & Monetary Interest & \textit{"reduce the credibility of the whole system"} [p05] & 2 \\
        & Response Type & \textit{"organic to read"} [p07], \textit{"a little confused at first"} [p01]& 1 \\
        & Explanations & \textit{"to judge how truthful the information is, I didn't find it that helpful"} [p04] & 0 \\
        & Replies & \textit{"didn't notice ChatGPT pointing out that I could ask [a follow up]"} [p05] & 0 \\\midrule
        
        \multirow{5}{*}{\rotatebox{90}{\textbf{Group 2}}} & Ordinal CS & \textit{"immediately understandable without having to read a lot of text" }[p02] & 0 \\
        & Direct Quotes & \textit{"one can see right where it is written"} [p07] & 1 \\
        & Disclaimer & \textit{"it doesn't help me, I know [...] I have to be careful there"} [p06] & 0 \\
        & Phrasing & \textit{"had to look very carefully to see if it was true"} [p08] & 1 \\
        & Confidence Threshold & \textit{"is more of a toy"} [p04],  \textit{"choose how creative or safe the AI should be"} [p05] & 5 \\\midrule
        
        \multirow{7}{*}{\rotatebox{90}{\textbf{Group 3}}} & Metric CS & \textit{"what it's based on, that [the score] is 70\%"} [p03] & 2 \\
        & Source Links (Pop-Up) & \textit{"it's easier than if I have to scroll around there"} [p06] & 1 \\
        & Source Quantity & \textit{"only [...] how many sources [...] not as good as these source links"} [p04] & 0\\
        & Political Spectrum & \textit{"just because it has [a political orientation] doesn't mean it's wrong"} [p04] & 0\\
        & Ethical \& Legal Consid. &  \textit{"didn't feel like [the features] were helping"} [p03] & 0 \\
        & Color & \textit{"it would have been enough if it was just highlighted in red"} [p03] & 2 \\
        & Drill-Down Dashboard &  \textit{"which of the three [tabs] is particularly important to focus on now"} [p05] & 0\\
        \bottomrule
    \end{tabularx}
    \caption{Assessment of each instantiation in the three prototypes with exemplary quotes and selections as top features.}
    \label{tab:woz_quotes}
\end{table*}

The participants were on average 35.11 years old (22 to 59 years) and a gender distribution of 4 women and 5 men. We included a broad range of participants stemming from different professions (e.g., students and retirees) and with different proficiency of using LLMs. During the interaction with the prototypes, we encouraged our participants to explain their thoughts and behavior (i.e., think-aloud). The following post-hoc interviews took on average 12 minutes and 32 seconds. Both parts were transcribed automatically and evaluated with a thematic analysis \cite{braun_using_2006}. When analyzing the sessions, we identified 470 text passages and 89 codes.

\paragraph{First Interface.} The interaction with the first interface lasted on average 6 minutes and 48 seconds. Six of the nine participants first noticed the \textit{colored ordinal CS} that was perceived as \textit{"easy to interpret"} [Participant 1 (p01)] and \textit{"immediately understandable without having to read a lot of text"} [p02]. One participant explicitly liked the color scheme, as shown in \autoref{tab:woz_quotes}. Despite being considered a top feature by three participants, others struggled with the composition of the CS. The participants assumed \textit{"that it is composed of the number of sources, the quality of the sources, [and] the number of uncertain text passages"} [p09] and they suggested adding information explaining how the score is calculated.

Despite being overlooked twice, the button displaying \textit{source links} was included six times as a top feature because it was \textit{"one of the most helpful"} [p05] features and enabled users to see \textit{"at a quick glance [...] where the information comes from"} [p04]. The participants suggested that clicking on \textit{source links} in the footnotes forwards to the source. Three of the nine participants voted \textit{source quality} into their cross-category top features, as it is \textit{"an easy indicator of how good the sources are"} [p01]. Our participants recommended an explanation so \textit{"it is clearly disclosed how to arrive at such source quality"} [p03]. Another suggestion was to evaluate each source to determine \textit{"which source is highly objective"} [p05].

A feature strongly discussed was \textit{monetary interest} which was \textit{"super striking"} [p02] and, therefore, included as a top feature twice. Five participants lost confidence not only in the marked sentences but in the entire response if paid content occurred. They indicated that this \textit{"is not an answer [they] want to work with"} [p02] and that it would \textit{"reduce the credibility of the whole system"} [p05]. 

One participant included \textit{response type} as a top feature because the underlined text indicated uncertainty, but maintained an \textit{"organic to read"} [p07] representation. Two other participants were \textit{"a little confused at first"} [p01] thinking \textit{"that [the underlined text] was a link"} [p01]. Additionally, four participants had trouble understanding the button for "uncertain information". These assessments emphasize the relevance of this feature but allow for discussion about instantiations. Two features (\textit{explanations} and \textit{replies}) received only little attention during the WOz sessions. Only three participants found \textit{replies} as a feature in the interface. Two participants were even surprised that it was listed as an extra feature since \textit{"of course, you can ask follow-up questions normally in this system at any time"} [p04].

\paragraph{Second Interface.} The interaction with the second interface lasted on average 6 minutes and 15 seconds. All participants had to interact with the \textit{disclaimer} first, but already knew the stated information (e.g., "I already know this" [p02] or "this is clear to me" [p07]). Seven participants stated that the disclaimer served no purpose while two participants found the disclaimer useful before the first interaction and \textit{"otherwise [...] rather obsolete"} [p05]. 

In five cases the \textit{confidence threshold} was clicked directly after the disclaimer. The post-hoc interview revealed that three participants thought \textit{"the threshold is more of a toy"} [p04] and that they \textit{"find it exciting [...] how the answers change"} [p06]. Nevertheless, five participants listed \textit{confidence threshold} as one of their top features because they \textit{"can choose how creative or safe the AI should be"} [p05]. One participant suggested that \textit{"the AI could tell from the question [...] which confidence is appropriate"} [p05]. The participants paid less attention to the \textit{ordinal CS} compared to the colored variant and described the same issues of understanding the calculation. Compared to the \textit{colored ordinal CS}, the variant was never included as a top feature.

\textit{Direct quotes} were included by one participant as top feature because it showed \textit{"exactly how it is written"} [p07] in the source, and they could \textit{"see right where it is written"} [p07], which made it easier for them to check the facts. 
Five participants criticized that the \textit{phrasing} feature was \textit{"not clearly marked"} [p03] resulting in two participants overlooking it. One participant chose the feature in his top features because different phrasing \textit{"gets a special weight"} [p02] for the truth of the answer. 

\paragraph{Third Interface.} Interaction with the third interface was the shortest, lasting on average for 5 minutes and 42 seconds. 
Three participants investigated the \textit{drill-down dashboard} first. The dashboard was rarely perceived as a single feature but generally viewed and evaluated in relation to the features within resulting in only six text passages. One participant suggested that \textit{"one could highlight [...] which of the three [tabs] is particularly important to focus on right now"} [p05].
Five participants used the \textit{metric CS} immediately after the dashboard and two included it as a top feature. The feature \textit{"graded a little bit better than Low, Medium, High"} [p09] but again the composition of the score \textit{"what it's based on, that [the score] is 70\%"} [p03] was unclear. Therefore, a clickable informative text was proposed.

Three participants found the \textit{political spectrum} \textit{"a bit out of place"} [p03] in a medical question, as it made \textit{"no sense with a topic like this"} [p05] but appreciated it \textit{"when it comes to something political"} [p07]. Participants indicated that the feature does not \textit{"assist them in determining the truthfulness"} [p03] because if the answer \textit{"doesn't fit my political views, I don't like the answer"} [p03], although the response might be correct. Three participants perceived \textit{ethical} and five participants perceived \textit{legal considerations} as \textit{"good and important"} [p01] but would refrain from using these features since \textit{"you don't read through all of [the text]"} [p05] and \textit{"didn't feel like [the features] were helping"} [p03].
Compared to the drill-down display from the first interface, the pop-up instantiation of the \textit{source links} was only selected once as a top feature because \textit{"it's easier than if I have to scroll around there"} [p06] and \textit{"it does not hide the actual text"} [p04]. The \textit{source quantity} was only addressed twice. The number of sources was expected to compensate for the quality of the sources \textit{"since [ChatGPT] is confident [...], even if the sources individually [...] are not that good"} [p09] and \textit{"that might be because he found relatively many sources"} [p09]. 

Three participants were confused by the \textit{colored} display of underlined text because it looked like \textit{"a typo"} [p01] or \textit{"thought it was kind of a link"} [p03]. Two participants chose it as their top feature because it was \textit{"clearly marked [...] from the beginning"} [p03]. \textit{Color} was also well received in conjunction with other features across the interfaces. The \textit{colored ordinal CS} performed significantly better than its gray variant, and the colored scaling of the \textit{source quality} was also positively highlighted. In terms of underlining and highlighting uncertain information, \textit{color} emerged from the interviews as a favorite.
Participants suggested varying the feature selection depending on the question type, so \textit{"the AI can tell based on the question what features it needs to display"} [p05] and a note for first-time users to display the features. Another suggestion was to combine \textit{replies} with \textit{response type}, that \textit{"if the information is uncertain, you should ask"} [p05] a follow-up question.

\subsection{Quantitative Prototype Evaluation}

All nine participants answered the questionnaires and followed the best-worst scaling approach. In the following, we first present the usability of the system. Second, we present the feature importance based on the rating of the participants. 

\begin{table*}[t]
    \centering
    \begin{tabular}{@{}llrlrlrlrlr@{}}
\toprule
\textbf{}    & \multicolumn{2}{c}{\textbf{Confidence Score}} & \multicolumn{2}{c}{\textbf{Sources}} & \multicolumn{2}{c}{\textbf{Disclosure}} & \multicolumn{2}{c}{\textbf{Vis. Aid \& In Text}} & \multicolumn{2}{c}{\textbf{Customization}} \\ \midrule
\textbf{1st} & \textit{Src. Qual.}       & \textit{(8)}      & Drill-Down        & (12)        & Mon. Inter.         & (15)         & Color                      & (14)                & Conf. Thr.         & (17)             \\
\textbf{2nd} & Metric CS                 & (7)               & Src. Qual.             & (4)         & Legal Con.            & (3)          & Resp. Type                 & (8)                 & Dashboard               & (-4)             \\
\textbf{3rd} & Colored Ord. CS           & (6)               & Dir. Quotes            & (1)         & Ethical Con.          & (-1)         & Explanations               & (-5)                & \textit{Resp. Type}     & \textit{(-5)}    \\
\textbf{4th} & Ordinal CS                & (-21)             & Pop-Up            & (-2)        & Disclaimer               & (-7)         & Phrasing                   & (-17)               & Replies                 & (-8)             \\
\textbf{5th} & -                         &                   & Src. Quant.            & (-15)       & Pol. Spectr.          & (-10)        & -                          &                     & -                       &                  \\ \bottomrule
\end{tabular}
    \caption{Results of the best-worst scaling per category with the score included in brackets. Features in italics were included to allow performing best-worst scaling but do not belong to the category.}
    \label{tab:best-worst}
\end{table*}

\paragraph{Usability.} To assess the usability of the prototypes, we asked them to assess ten statements of the system usability scale (SUS) \cite{brooke1996sus} on a 5-point Likert scale (ranging from 1 - "totally disagree" to 5 - "totally agree").
Our participants perceived the first prototype as easy to use (average: 4.11, standard deviation: 0.58) and stated they would frequently use the system (4.00, 1.07). Drawbacks revolve around the consistency of the features, where the participants were indifferent to the statement that "there were too many inconsistencies in the system" (2.44, 1.25).

The same problem holds for the second prototype where the participants perceived slightly fewer inconsistencies (2.11, 1.46). They would also like to interact with this system frequently (4.67, 0.38) and strongly disagree with the statement the system is "unnecessarily complex" (1.22, 0.38). Despite these promising assessments, the participants also do not feel fully confident when using the system (3.56, 0.79), the lowest score out of the three prototypes.

The third prototype got the lowest score for the likelihood of frequent use (3.67, 1.11) and ranked last regarding its complexity (1.89, 0.90). The prototype showed especially few inconsistencies (1.67, 0.69) and a quick learning to use this system (4.22, 0.82).

\paragraph{Feature Importance.} Second, we conducted a best-worst scaling for each feature category \cite{Finn1992} which has been recommended to rate feature-based benefit importance over other comparisons due to its easy implementation and "easy to explain" results \cite{cohen2003maximum}. Evaluating all 20 instantiations of the features in groups of four would create an enormous amount of configurations. Therefore, we conduct a best-worst scaling to the instantiations within one category. For each feature category, we ask our participants to select the best and worst feature instantiation for multiple feature combinations. The combinations were created based on the balanced incomplete block design \cite{bose1939construction}. We grouped categories with only two features (i.e., visual aid and in text) together, since the comparison would be binary otherwise. For categories containing three instantiations (i.e., Confidence Score and Customization) we included one additional feature which we ignored during the analysis. Each instantiation's rating is determined by adding 1 to the score for each assessment as a best instantiation and by subtracting 1 as a worst instantiation \cite{Finn1992}. An overview of the results of the best-worst scaling is shown in \autoref{tab:best-worst}.

When assessing the best instantiation of the CS, we see a slight tendency toward the metric CS (score = 7), just before a colored ordinal CS (6). In this group of feature instantiations, the included source quality was perceived as even more essential to include (8). However, for the sources category, source quality only ranks second (4), just behind the provision of source links in a drill-down menu (12). This representation is strongly preferred to the instantiation as a pop-up window (-2) which is in line with the interview results. In the disclosure category, monetary interests (15) were perceived as the best feature to include. This might be due to the impact of paid content on the acceptance of the response specified in the interviews. A disclaimer (-7) and the political spectrum (-10) were perceived as the worst features in the disclosure category to include. In the combined visual aid and in text category, the visual features (color, 14 and response type, 8) greatly stood out compared to the textual features (explanations, -5 and phrasing, -17). In the customization step, the confidence threshold (17) was the best-rated feature with 21 points more than the second-rated drill-down dashboard (-4).

\section{Study 2: Artifact Generation}
\label{sec:artifact}

Based on the feature prioritization of the WOz and our prototypes, we developed the "Hallucination Identifier for Large Language Models" (HILL) that should enable users to identify potential hallucinations.

\subsection{Artifact Development}

The existing prototypes provided several interface features to build on. With these interfaces at hand, we derived some technical requirements based on existing guidelines \cite{tomitza2023minimum} and following iterative discussions. Based on the requirements to trust text-based generative AI, usage should be as intuitive as possible \cite{tomitza2023minimum}. Therefore, we decided to build a web application that runs independently of the operating system, so the end users are not required to install or update any service. We implemented our frontend, which covers all interaction with the users, based on the JavaScript Framework Vue.js \cite{vuejs2014}. The backend, which ensures the communication between the APIs of ChatGPT and HILL as well as the calculation of the CSs and other measures, is built with Express.js\footnote{\url{https://expressjs.com/}}. OpenAI provides an easily accessible API to ChatGPT \cite{openai_Api}, therefore, we connect HILL to ChatGPT instead of other established LLMs like BARD or LaMDA. This fulfills the requirements of easy integration with existing resources and enables easy transferability to other AI models \cite{tomitza2023minimum}.

\paragraph{Frontend Development} The \textit{frontend} of HILL consists of four areas which can be seen in \autoref{fig:frontend_prototype}. First, we built upon the comments of the WOz participants and explained the included features in a text box positioned at the top of the application site (area A). The text box contains explanations of the features \textit{confidence score}, \textit{political spectrum}, and \textit{monetary interest} as well as two additional explanations for \textit{hallucinations} and \textit{self-assessment score}. The second area B is a chat interface with messages between the LLM and the user. We used the same icons as OpenAI for ChatGPT and unknown users \cite{openaiWhatDoes}. We provided three extensions below the generated response, from left to right, a button with "+Sources", a metric CS, and a "+More details" button. The last button opens the third area C of an additional drill-down dashboard that contains more detailed information regarding the assessment of the political spectrum, the monetary interest, hallucinations, and the self-assessment score. Despite having only mediocre results in the best-worst scaling, we include these features in a drill-down dashboard to allow for additional information only if desired to avoid overburdening the users. A similar dashboard opens below the response when clicking on the left "+Source" button and shows a list of the sources supporting the generated response. We used the dashboard option rather than the pop-up instantiation based on the best-worst scaling. The fourth area D is shown at the bottom of the interface where users regenerate the response of the model and can type in additional messages.

\paragraph{Backend Development} The \textit{backend} enables communication with OpenAI's API as well as the calculation of the CS. During the response generation with HILL, we conduct multiple calls of the API. First, the user input is posed to ChatGPT-v3.5 creating the initial response that serves as the basis for further analysis. This mimics conventional communication with LLMs where users can ask follow-up questions based on previously generated responses. These follow-up questions were one feature developed by \cite{leiser_chatgpt_2023}.

As a next step, HILL follows with a second request to ChatGPT to identify uniform resource locators (URLs) for sources supporting the generated response for the user input. Therefore, we include the user input as well as the initial response in our request to OpenAI's API. The prompts for all additional requests follow OpenAI's best practices for prompt engineering and can be found in our supplement \cite{openaiBestPractices}. We set the temperature of this request to \texttt{0} since only factual responses are desired. To ensure source validity and further increase the benefits for users \cite{tomitza2023minimum}, we additionally verify the URLs provided by ChatGPT by calling the URL and only presenting the source to the user if the URL returns the status code \texttt{"200"} indicating a functioning URL. In our initial trials, this validation removed about half of the ChatGPT-generated and -hallucinated sources. Validated sources are presented at the bottom of the screen in the source menu.

Following the source identification, we, third, ask ChatGPT to assess other ethical considerations in the initial response regarding the degree of monetary interest or political opinion. Monetary interest has been shown to be the most relevant disclosure feature in the feature prioritization conducted above. We ask the LLM to self-assess a score for both dimensions on a Likert scale. Self-assessment of LLM responses has already been conducted in previous studies to ensure the validity of the response \cite{manakul2023selfcheckgpt}. The degree of monetary interest is provided on a Likert scale ranging from $0$ ("Very unlikely to be paid content") to $10$ ("Very likely paid content"). For the political spectrum, the response is given on a scale between $-10$ and $10$ where $-10$ stands for "extreme left political spectrum", $0$ is "neutral", and $10$ is "extreme right political spectrum". Both score assessments are briefly explained as well. The scores and the explanations are shown in the drill-down dashboard on the right side of the interface.

For the fourth request, we ask ChatGPT to identify potential misinformation based on the response type. In our request, we encourage the LLM to "act as an independent fact checker" with the given question and output. The model should then identify all "errors in factual information and subjective statements" in the response text to ensure and fulfill the requirement of response neutrality \cite{tomitza2023minimum}. The prompt specifies that the response is a JSON object including the number of errors and subjective statements as well as quotes and explanations on why the LLM considers them to be errors. These errors are then highlighted in the initial response in the user interface and the explanations given on the right side in the drill-down dashboard. 

To validate the initial response, we send a last request to the OpenAI API which should "assess the accuracy of the given answer to the given question/task and provide a self-assessment score between $0$ and $100$ where $0$ is 'totally factually wrong' and $100$ is 'totally factually true'". Again this score is returned as JSON-object with the score and a short explanation of the assessment.

Based on these additional requests, we compute an overall \textit{confidence score} for the reply. The CS is a weighted aggregation of the scores returned by the additional requests. To achieve proper comparison, we scaled the self-assessment score, political spectrum, and monetary interest all to a normalized scale between $0$ and $1$. We further transformed the provided sources and the response type into a score between $0$ and $1$. For the source score, we assumed that if a functioning URL is returned, this already significantly boosts the confidence in the response. Therefore, we determine the normalized source score $norm_S = 0.5 + 0.1*(number\, of\, sources-1)$. This means, that the first validated source sets the source score to $0.5$ while every additional validated source further increases the score by $0.1$. If no sources are provided, $norm_S$ is set to $0$, if more than 6 sources are provided, $norm_S$ is $1$. We determined the normalized score for the response type $norm_T$ in a similar fashion. We assumed, that if no hallucinations are found $norm_T$ is set to $1$, and if 4 or more hallucinations are identified $norm_T$ is set to $0$. This lead us to $norm_T = 1 - (number\, of\, hallucinations) / 4$ for all other values. Taking all normalized scores into account we determine the overall CS of the initial response with 
\begin{eqnarray*}
confidence\, score &=& 0.1 * norm_S + 0.5 * norm_{CS} + 0.05 * norm_{P}\\
&& +\,0.05 * norm_{M} + 0.3 * norm_{T}
\end{eqnarray*}

To sum up, we use LLMs to evaluate LLM output. An alternative could be to use external evaluation as suggested in literature \cite[e.g.,][]{galitsky2023truth,peng2023check,gao2023rarr}. We purposefully chose another route for various reasons. First, this approach is easy to implement, since it relies on one API for all requests. Second, our approach is LLM-model agnostic. While we base our model on ChatGPT-v3.5, promising results might be also transferred to other LLM models. Nonetheless, we acknowledge other methods, such as external evaluation as other promising approaches.

\subsection{Artifact Evaluation}

To evaluate the interface of HILL, we conducted a brief online survey with 17 participants. We additionally validated the identification of hallucinations based on the second version of the Stanford Question Answering Dataset (SQuAD 2.0) \cite{squad_2018_rajpurkar}. Furthermore, we conducted five interviews to assess the influence of HILL on user reliance.

\paragraph{Online Survey} For the interface evaluation, we posed five questions in total with three questions regarding the education of the users and two regarding the clarity of the user interface. Each question was answered on a 7-point Likert scale ranging from 1 ("strongly disagree") to 7 ("strongly agree"). The questions were framed in German language to ease the survey participation for the respondents. We show English translations in \autoref{tab:proto_eval_questions}. 

\begin{table*}[tbp!]
    \centering
    \begin{tabularx}{\textwidth}{lXrrr}
         \toprule
         \textbf{ID} & \textbf{Phrasing} & \textbf{ChatGPT} & \textbf{HILL} & \textbf{Improvement}  \\\midrule
         \textbf{E1} & I consider the validity of \textit{<product>} to be high. & 4.82 (1.44) & 6.00 (0.59) & 24.39\% ***\\
         \textbf{E2} & The \textit{<product>} user interface draws attention to the fact that hallucinations may be included in the responses. & 3.65 (2.11) & 6.71 (0.21) & 83.37\% *** \\
         \textbf{E3} & I see statements that are passed off as facts by \textit{<product>} as critical. & 5.12 (1.99) & 3.76 (1.83) & 26.44\% ** \,\\
         \textbf{UI1} & I perceive the user interface of \textit{<product>} as concise. & 6.35 (0.70) & 6.24 (0.65) & $-1.85\%\ ^{ns}$\ \\
         \textbf{UI2} & The user interface of \textit{<product>} offers me all the features relevant for use. & 5.82 (0.85) & 6.12 (0.81) & $5.05\%\ ^{ns}$\ \\\bottomrule
    \end{tabularx}
    \caption{Results of the survey regarding the usability of ChatGPT compared to our HILL artifact. The shown scores are the average of the replies on a Likert scale from 1 ("strongly disagree") to 7 ("strongly agree"), standard deviation in brackets.} 
    \label{tab:proto_eval_questions}
\end{table*}

Regarding the education of the users, we see that all three statements have significant improvement when using HILL. The 17 participants considered the validity of HILL to be 24.39\% higher than the validity of ChatGPT (6.00 vs. 4.82, $p < 0.0001$) and that the HILL user interface better draws attention to potential hallucinations than the ChatGPT interface (6.71 vs. 3.65, $p < 0.0001$). In the third statement, the participants answered they perceive factual statements of HILL as less critical compared to ChatGPT (3.76 vs. 5.12, $p = 0.008$). While this indicates an increased trust in the system, it also exposes an increased risk of overreliance if HILL misses hallucinations.

We also posed two questions about the user interface and asked whether the interface was perceived as concise and if the interface offered all relevant features. Both questions had similar replies among our participants and we identified no significant difference between ChatGPT and HILL. The interface of HILL was perceived as slightly less concise than the ChatGPT interface (-1.85\%) which stems from the additional features incorporated in this artifact.

We posed a final question where we asked our participants whether the validation of statements containing alleged facts is easier with HILL. Out of the 17 participants, 12 "strongly agreed" and 4 "agreed" with this statement. We acknowledge a potential social desirability bias here which needs to be removed and validated in future blind evaluations of HILL compared to the traditional LLM interfaces.

\paragraph{Performance Validation} Besides the questionnaires related to usability, we also analyzed the performance of HILL in the identification of hallucinations. For that, we used SQuAD 2.0 \cite{squad_2018_rajpurkar} which contains about 150,000 questions. Each question is associated with a short text prompt containing information regarding different categories like \textit{prime numbers}, \textit{Huguenots}, or \textit{1973 oil crisis}. The data set differs between 100,000 answerable questions and 50,000 unanswerable questions regarding the provided texts.

For our evaluation, we posed 64 answerable and 64 unanswerable questions to HILL, four randomly picked each from 16 different categories. An exemplary "answerable" question for the category \textit{prime numbers} would be "Which theorem would be invalid if the number 1 were considered prime?" (Answer: Euclid's fundamental theorem of arithmetic). These questions are factually true and easy to validate. Unanswerable questions included negations that are difficult to detect or factually wrong assumptions in their question, like "Who included 1 as the first prime number in the mid-20th century?". Assessment of HILL's responses to unanswerable questions was challenging since the LLM might have learned the information during model training sessions. Therefore, we manually validated whether factual responses were available and assessed the reply as "correct" if the artifact's response either correctly stated they did not know the answer or the correct factual answer was given. Based on these questions, we analyzed whether HILL identified hallucinations or not and whether the given answer was correct or not. We, additionally,  gathered all available scores of HILL including the self-assessment score and the overall CS. \autoref{tab:confusion_matrix} shows an overview of the totals.

As seen in \autoref{tab:confusion_matrix}, HILL generally identifies hallucinations in wrong answers and identifies no hallucinations in factually correct answers. Overall, it achieves an accuracy of 70.3\% for answerable questions and 64.0\% for unanswerable questions. In our evaluation, when HILL produced wrong answers, it mostly identified hallucinations (recall of 60.0\% for answerable and 66.7\% for unanswerable questions). One drawback is that the identification of hallucinations does not necessarily mean the answer is wrong (precision of 52.2\% for knowable, 41.4\% for unanswerable questions).

Comparing the responses to answerable and unanswerable questions, we recognize that HILL identified hallucinations more frequently in responses to unanswerable questions compared to answerable ones (44 hallucinations in 29 responses vs. 29 hallucinations in 23 responses). Since HILL needed to answer questions without evident factual responses, we suspected more hallucinations in these responses and found more. However, for unanswerable questions, HILL identified more hallucinations on correct answers than on wrong answers (26.6\% vs. 18.8\%). One of the reasons might be that the identified hallucinations were considered 
subjective statements in 19 responses compared to only 5 responses to answerable questions. These subjective statements were frequently considered as subjective because they provided no references and sources supporting the claim.

Analyzing the CS and self-assessment score for each question yielded no interesting results. HILL generally ranked responses where no hallucinations were identified as higher (85.44 vs. 80.79 in CS) as well as responses where the correct answer was given as higher compared to incorrect answers given (83.80 vs. 82.44 in CS).

\paragraph{User Interviews} We additionally conducted semi-structured interviews to gain a deeper understanding of users' reliance on HILL. In each interview, we presented the users with four questions from the SQuAD 2.0 dataset \cite{squad_2018_rajpurkar} first showing an interaction with ChatGPT and afterward an interaction with HILL. These interactions were provided via videos to follow a standardized procedure. Two of the presented questions showed general knowledge questions (divisors of prime numbers, location of a large European river) as well as two questions with higher knowledge requirements (second most abundant element, pre-allocation of resources in packet-switching). We recorded and transcribed the interviews verbatim \footnote{We encountered technical difficulties in one interview and therefore only used the notes taken during the interview.}.

In total, we interviewed five participants (4 males, 1 female) based on convenience sampling. Interviews were coded in an open coding approach with a focus on the design features of HILL. The participants all stated they frequently use LLMs for personal as well as professional reasons. Most commonly, they used LLMs for text summaries and recaps of unfamiliar concepts. We first present the results of the general interaction with HILL. Then we dive deeper into certain features of HILL where we found interesting patterns in the statements of the participants. 


When asking questions to LLM responses, the participants generally assess if the response \textit{"fits into [their] current understanding} [i01] and \textit{"sounds logical"} [i04]. If that is the case, they would rely on the response (i01, i04). This opens up the case of confirmation biases, where humans only look for desired answers, therefore, highlighting the importance of mitigating overreliance on LLMs. This reduction can be achieved by some of the HILL design features presented below. 

The confidence score was seen as the \textit{"most important"} [i01] design feature by many of the participants. However, there were also some challenges in grasping the meaning behind it. For example, one participant would need some time to adapt their reliance and \textit{"get a feeling"} [i03] of the performance in order to rely on the confidence score. Further, understanding the meaning of the numerical value might be a challenge, as \textit{"a confidence score of 77\% is too little to rely on"} [i03] on the model. The interviewees additionally highlighted the benefits of the self-assessed score since they \textit{"can better grasp the basic points"} [i03] with the provided explanation. Further, sources were seen as \textit{"particularly helpful"} [i02] and the \textit{"most important and helpful"} [i03] feature. However, the question of source quality arose. One of the sources included Wikipedia, which was critically questioned by some participants as \textit{"not the most reliable source"} [i03 and i04]. At the same time, just having sources, even Wikipedia, \textit{"increases confidence"} [i03], as they would expect, that \textit{"the content of Wikipedia was used and the information was simply extracted from it"} [i03]. This highlights, that source quality might be a subjective feature, as some participants perceive the same source with varying quality.

HILL also directly states identified hallucinations. If that is the case, participants would \textit{"first look at hallucinations, then at the sources to assess whether it is correct what the response states"} [i04]. Further, only hallucinations are presented, participants \textit{"would follow up"} [i02] on that, highlighting the importance of presenting the hallucinations. Especially in tasks where they are inexperienced, the participants stated they \textit{"would not reassess the response of ChatGPT, since they have a higher trust in ChatGPT"} [i01]. However, \textit{"now when the confidence is shown and it is small, [they would] reassess the response"} [i01]. Other interviewees stated they rather use ChatGPT \textit{"if is about general questions and only the general information is important"} [i02]. While monetary interest and political spectrum got little attention from many participants, one participant particularly appreciated the features as these can be used to \textit{"influence the formation of opinions"}[i05], but also acknowledged that this might be \textit{"difficult to measure"}[i05].

Overall, HILL was perceived as positive and participants made clear that the features have the potential to reduce blindly following the LLM-generated response, i.e., overreliance. Interestingly, one participant stated when doubting the answers of ChatGPT they would also \textit{"open a new ChatGPT window and ask the same question again to compare the answers and see whether the same answer comes out twice or whether the answer is different"} [i01]. This shows that the design HILL follows a similar approach to the decision-making of some humans: validating LLMs by using LLMs.

\begin{table*}[tbp!]
    \centering
    \begin{tabular}{lcccc}
         \toprule
         & \multicolumn{2}{c}{\textbf{"Answerable" Questions}} & \multicolumn{2}{c}{\textbf{"Unanswerable" Questions}} \\\midrule
         \textbf{Answer} & Correct & Wrong & Correct & Wrong \\\midrule
         \textbf{No Hallucinations Found} & 33 & 8 & 29 & 6 \\
         \textbf{Hallucinations Found} & 11 & 12 & 17 & 12 \\\bottomrule
    \end{tabular}
    \caption{Confusion matrix of the identification of hallucinations of our prototype on questions of SQuAD 2.0 \cite{squad_2018_rajpurkar}}
    \label{tab:confusion_matrix}
\end{table*}

\section{Discussion}

With the results of the WOz and the following development of the HILL artifact, we discuss in this section the contribution of both studies as well as the limitations and provide possible directions for future improvements.

\subsection{Contributions}
In this study, we followed a user-centered approach to propose the novel artifact HILL, the "Hallucination Identifier for Large Language Models". We identified the design features using a WOz study and on this basis implemented HILL. We put a specific focus on user needs and perceptions as a central aspect of our study. While current research aims at improving performance metrics of LLMs or correcting LLM output \cite[e.g.,][]{galitsky2023truth,peng2023check,gao2023rarr}, we acknowledge that the systems will not be entirely hallucination-free in the near future. Therefore, we identify design features that enable users to detect and act upon hallucinations. We see this as a major gap in current research and aim to address this with our study and the resulting artifact HILL. We provide a detailed description of the procedure identifying relevant design features to be included in HILL as well as its underlying architecture. The evaluation of HILL shows promising results regarding its potential to support users in identifying hallucinations. With that, our study can serve as a guideline for the future development of similar artifacts incorporating user feedback when developing interfaces to detect hallucinations or other issues in AI model responses. In the following, we discuss our contributions in more detail.

Building on previously proposed features \cite{leiser_chatgpt_2023}, we first implemented three prototypes in Figma. Features that were perceived as especially promising in previous work had different instantiations to allow for different representations. Nine users assessed the prototypes and the benefits of each instantiation. In the following evaluation regarding the usability of the system and best-worst scaling, we evaluated the best instantiations for each of the six categories. We found, that the users liked a metric CS more than other presentations and the sources should be included as links in a drill-down feature. Additionally, visual highlighting by color and underlining different response types were perceived as more helpful than textual changes in the response phrasing. Users found monetary interest in the response to be essential since they would not trust the response if it included paid content. Users liked a confidence threshold to ensure relevant and factually true responses.

While taking into account user feedback of the WOz, we developed HILL which builds on the existing API of ChatGPT by OpenAI. With that, HILL is a complementary approach to existing LLMs rather than a competing one which can be simply included as an additional interface. We included the features metric CS and color as features performing best in two categories in the best-worst-scaling (CS, Visual Aid \& In Text) as well as two drill-down dashboards. The first dashboard contains source links to references, and the second drill-down dashboard has additional features about disclosure like monetary interest and the response type. These features are assessed by additional requests to the ChatGPT API using the initial request and the model's generated response. All these features are incorporated in an overall CS by weighing the self-assessment score and the presence of potential hallucinations the most. In a following evaluation survey, 17 participants saw the interface as beneficial to assess potential hallucinations in LLM responses while maintaining an easy-to-use interface.

We additionally evaluated the functionality of HILL by posing questions from SQUAD 2.0 \cite{squad_2018_rajpurkar}. With 128 questions, we saw that HILL can identify hallucinations. In initially wrong answers of LLMs, we found hallucinations in 24 out of 38 cases. This indicates that our HILL artifact is able to detect artificially generated information in LLM responses and, therefore, helps users assess the factual accuracy of LLM responses. With correct responses, our model might still identify hallucinations in the form of subjective statements that need further assessment by the users. We also saw that users perceived statements without identified hallucinations in HILL as less critical than the same statements in ChatGPT. This perception was especially driven by the presence of sources and the indication of confidence scores which provide a convenient starting point for users to manually assess the model's response. This showed slight tendencies toward an increased reliance on HILL. Since HILL is unable to identify all hallucinations, this could also increase overreliance on factually wrong answers. Future endeavors should investigate how to avoid this increased overreliance as well as further possibilities to improve the accuracy of HILL and enhance the quality of the provided sources. Therefore, we encourage users to use HILL to identify hallucinations in LLM responses but also rely on other resources to conclude factually correct answers.

By following a user-centered approach, we identify important design features proposed by users. Taking into account user feedback is essential to enable frequent use of LLMs and AI in general. Therefore, prioritizing features based on a WOz can guide the development of future development of AI artifacts in general. Wherever features desired or frequently relied upon by users are available, a prioritization of these features might lead to artifacts with easier use. Therefore, designers and developers can build on our approach to develop their own user-centered AI artifacts.

Further, our findings might generalize beyond question and answering tasks. In our interviews, we saw a frequent use of LLMs for validating and summarizing unknown concepts where similar features might be helpful \cite{maynez2020faithfulness}. Abstractive summaries are especially prone to hallucinations. Therefore, our proposed design features could help enable users to detect hallucinations. While in question answering the features can be built for the full answer, a more fine granular partitioning might be appropriate for summarization, such as the sentence or paragraph level. Similarly, for machine translation, hallucinations can sometimes be found \cite{raunak2021curious}. Additionally, even beyond NLP, e.g., hallucinations in computer vision or image creation occur \cite{zhou2023analyzing}. While our features are developed for text-based hallucinations, some features might also be adequate beyond text-based problems, such as confidence scores or disclaimers. Overall, our study addresses the gap in how to enable users to detect and act upon hallucinations from a user-centered approach. 

\subsection{Limitations \& Future Research}

Our study, however, is not without limitations. First, we only provide a preliminary evaluation of our HILL artifact. While assessing the user interface and the functionality of the model, including user interviews, we did not quantitatively investigate the effects of identifying hallucinations on user behavior. Especially the limited sample size in user evaluation as well as assessing video instead of natural interactions invites future studies to investigate how the identification of hallucinations in LLM responses influences the user's interaction with the system and potential (over)reliance. Nonetheless, our study lays the groundwork for a more in-depth evaluation of HILL and similar artifacts.

Second, we developed only one instantiation for most of the proposed design features. While this enabled us to prioritize across features, it leaves the representation of each feature out of scope. In future studies, different instantiations might further improve the interaction with the system. Additional features might be included in the artifact as well as in the assessment of different features. This includes an instantiation of the confidence threshold which was not included in our artifact due to unclear technical implementation. Investigating different implementations of this feature could be a great first step to further improve HILL.

Finally, the confidence in the LLM response is currently self-assessed by additional requests to the ChatGPT API. This leads to increased communication between the client and the API requiring more network interaction as well as potentially higher pricing of the requests. One solution to reduce the communication effort could be to request this information only if the user desires it. The higher price, however, could be incorporated when pricing initial requests by OpenAI or other LLM providers. Another issue is that self-assessed scores rely on LLM responses and assessing these responses for potential hallucinations could result in recursively infinite requests. A better solution would be to allow a response assessment based on a search engine request. However, this would open other questions, especially regarding the source quality of search engine results.

\section{Conclusion}

In this study, we developed HILL, a Hallucination Identifier for Large Language Models. With the pressing issue of hallucinations in LLM responses, we followed a user-centered approach to avoid overreliance of users on factually incorrect responses. We prioritized the features to include in our artifact based on a WOz, where nine everyday users interacted with clickable prototypes. We found, that the inclusion of source links as well as a metric CS were of great importance to the users. Additionally, our participants heavily discussed the presence and highlighting of paid content in the response generation since that could \textit{"reduce the credibility of the whole system"} [p05]. Based on the feature prioritization we built a web-based artifact that complements, not competes, with other existing technical efforts to reduce hallucinations in LLMs. In a survey with 17 participants, we saw that the interface enables users to question the replies without overburdening them. We evaluated the detection of hallucinations on SQuAD 2.0 where we saw that HILL can correctly identify hallucinations in factually wrong responses. Further, the results of our user interviews highlight the potential of HILL to enable users to detect hallucinations. With this study, we contribute to improving LLMs and enabling future development of user-centered LLM-based systems.

\begin{acks}
We would like to thank Jakub Jeck, Benedikt Wolf, Nils Weber, and Marvin Voigt for aiding us in the development of the artifact. We additionally thank all the participants in our evaluation studies for their constructive and valuable feedback.
\end{acks}

\bibliographystyle{ACM-Reference-Format}
\bibliography{maintext}


\begin{thebibliography}{53}


\ifx \showCODEN    \undefined \def \showCODEN     #1{\unskip}     \fi
\ifx \showDOI      \undefined \def \showDOI       #1{#1}\fi
\ifx \showISBNx    \undefined \def \showISBNx     #1{\unskip}     \fi
\ifx \showISBNxiii \undefined \def \showISBNxiii  #1{\unskip}     \fi
\ifx \showISSN     \undefined \def \showISSN      #1{\unskip}     \fi
\ifx \showLCCN     \undefined \def \showLCCN      #1{\unskip}     \fi
\ifx \shownote     \undefined \def \shownote      #1{#1}          \fi
\ifx \showarticletitle \undefined \def \showarticletitle #1{#1}   \fi
\ifx \showURL      \undefined \def \showURL       {\relax}        \fi
\providecommand\bibfield[2]{#2}
\providecommand\bibinfo[2]{#2}
\providecommand\natexlab[1]{#1}
\providecommand\showeprint[2][]{arXiv:#2}

\bibitem[Bangor et~al\mbox{.}(2008)]%
        {bangor2008empirical}
\bibfield{author}{\bibinfo{person}{Aaron Bangor}, \bibinfo{person}{Philip~T Kortum}, {and} \bibinfo{person}{James~T Miller}.} \bibinfo{year}{2008}\natexlab{}.
\newblock \showarticletitle{An empirical evaluation of the system usability scale}.
\newblock \bibinfo{journal}{\emph{International Journal of Human-Computer Interaction}} \bibinfo{volume}{24}, \bibinfo{number}{6} (\bibinfo{year}{2008}), \bibinfo{pages}{574--594}.
\newblock


\bibitem[Bell(2023)]%
        {bell_fake_2023}
\bibfield{author}{\bibinfo{person}{Emily Bell}.} \bibinfo{year}{2023}\natexlab{}.
\newblock \bibinfo{title}{A fake news frenzy: why {ChatGPT} could be disastrous for truth in journalism}.
\newblock
\newblock
\showISSN{0261-3077}
\urldef\tempurl%
\url{https://www.theguardian.com/commentisfree/2023/mar/03/fake-news-chatgpt-truth-journalism-disinformation}
\showURL{%
\tempurl}


\bibitem[Borji(2023)]%
        {borji2023categorical}
\bibfield{author}{\bibinfo{person}{Ali Borji}.} \bibinfo{year}{2023}\natexlab{}.
\newblock \bibinfo{title}{A Categorical Archive of ChatGPT Failures}.
\newblock
\newblock


\bibitem[Bose(1939)]%
        {bose1939construction}
\bibfield{author}{\bibinfo{person}{Raj~Chandra Bose}.} \bibinfo{year}{1939}\natexlab{}.
\newblock \showarticletitle{On the construction of balanced incomplete block designs}.
\newblock \bibinfo{journal}{\emph{Annals of Eugenics}} \bibinfo{volume}{9}, \bibinfo{number}{4} (\bibinfo{year}{1939}), \bibinfo{pages}{353--399}.
\newblock


\bibitem[Braun and Clarke(2006)]%
        {braun_using_2006}
\bibfield{author}{\bibinfo{person}{Virginia Braun} {and} \bibinfo{person}{Victoria Clarke}.} \bibinfo{year}{2006}\natexlab{}.
\newblock \showarticletitle{Using thematic analysis in psychology}.
\newblock \bibinfo{journal}{\emph{Qualitative Research in Psychology}} \bibinfo{volume}{3}, \bibinfo{number}{2} (\bibinfo{date}{Jan.} \bibinfo{year}{2006}), \bibinfo{pages}{77--101}.
\newblock
\showISSN{1478-0887, 1478-0895}
\urldef\tempurl%
\url{https://doi.org/10.1191/1478088706qp063oa}
\showDOI{\tempurl}


\bibitem[Brereton(2023)]%
        {brereton_bing_2023}
\bibfield{author}{\bibinfo{person}{Dmitri Brereton}.} \bibinfo{year}{2023}\natexlab{}.
\newblock \bibinfo{title}{Bing {AI} {Can}'t {Be} {Trusted}}.
\newblock
\newblock
\urldef\tempurl%
\url{https://dkb.blog/p/bing-ai-cant-be-trusted}
\showURL{%
\tempurl}


\bibitem[Brockman et~al\mbox{.}(2020)]%
        {openai_Api}
\bibfield{author}{\bibinfo{person}{Greg Brockman}, \bibinfo{person}{Mira Murati}, \bibinfo{person}{Peter Welinder}, {and} \bibinfo{person}{OpenAI}.} \bibinfo{year}{2020}\natexlab{}.
\newblock \bibinfo{title}{{O}pen{A}{I} {A}{P}{I} --- openai.com}.
\newblock \bibinfo{howpublished}{\url{https://openai.com/blog/openai-api}}.
\newblock
\newblock
\shownote{[Accessed 25-08-2023]}.


\bibitem[Brooke(1996)]%
        {brooke1996sus}
\bibfield{author}{\bibinfo{person}{John Brooke}.} \bibinfo{year}{1996}\natexlab{}.
\newblock \showarticletitle{Sus: a “quick and dirty’usability}.
\newblock \bibinfo{journal}{\emph{Usability evaluation in industry}} \bibinfo{volume}{189}, \bibinfo{number}{3} (\bibinfo{year}{1996}), \bibinfo{pages}{189--194}.
\newblock


\bibitem[Carlini et~al\mbox{.}(2021)]%
        {carlini2021extracting}
\bibfield{author}{\bibinfo{person}{Nicholas Carlini}, \bibinfo{person}{Florian Tramer}, \bibinfo{person}{Eric Wallace}, \bibinfo{person}{Matthew Jagielski}, \bibinfo{person}{Ariel Herbert-Voss}, \bibinfo{person}{Katherine Lee}, \bibinfo{person}{Adam Roberts}, \bibinfo{person}{Tom~B Brown}, \bibinfo{person}{Dawn Song}, \bibinfo{person}{Ulfar Erlingsson}, {et~al\mbox{.}}} \bibinfo{year}{2021}\natexlab{}.
\newblock \showarticletitle{Extracting Training Data from Large Language Models.}. In \bibinfo{booktitle}{\emph{Proceedings of the 30th USENIX Security Symposium}}, Vol.~\bibinfo{volume}{6}. \bibinfo{publisher}{The USENIX Association}, \bibinfo{pages}{2633--2650}.
\newblock


\bibitem[Cohen et~al\mbox{.}(2003)]%
        {cohen2003maximum}
\bibfield{author}{\bibinfo{person}{Steve Cohen} {et~al\mbox{.}}} \bibinfo{year}{2003}\natexlab{}.
\newblock \showarticletitle{Maximum difference scaling: improved measures of importance and preference for segmentation}. In \bibinfo{booktitle}{\emph{Sawtooth Software Conference Proceedings}}, Vol.~\bibinfo{volume}{530}. Sawtooth Software, Inc. Fir St., Sequim, WA, \bibinfo{pages}{61--74}.
\newblock


\bibitem[Coulter and Bensinger(2023)]%
        {coulter_alphabet_2023}
\bibfield{author}{\bibinfo{person}{Martin Coulter} {and} \bibinfo{person}{Greg Bensinger}.} \bibinfo{year}{2023}\natexlab{}.
\newblock \showarticletitle{Alphabet shares dive after {Google} {AI} chatbot {Bard} flubs answer in ad}.
\newblock \bibinfo{journal}{\emph{Reuters}} (\bibinfo{date}{Feb.} \bibinfo{year}{2023}).
\newblock
\urldef\tempurl%
\url{https://www.reuters.com/technology/google-ai-chatbot-bard-offers-inaccurate-information-company-ad-2023-02-08/}
\showURL{%
\tempurl}


\bibitem[developers(2014)]%
        {vuejs2014}
\bibfield{author}{\bibinfo{person}{Vue.js developers}.} \bibinfo{year}{2014}\natexlab{}.
\newblock \bibinfo{booktitle}{\emph{Vue.js - The Progressive JavaScript Framework v3.0}}.
\newblock
\urldef\tempurl%
\url{https://vuejs.org/guide/introduction.html}
\showURL{%
\tempurl}
\newblock
\shownote{Accessed: 2023-08-25}.


\bibitem[Dow et~al\mbox{.}(2005)]%
        {dow2005wizard}
\bibfield{author}{\bibinfo{person}{Steven Dow}, \bibinfo{person}{Blair MacIntyre}, \bibinfo{person}{Jaemin Lee}, \bibinfo{person}{Christopher Oezbek}, \bibinfo{person}{Jay~David Bolter}, {and} \bibinfo{person}{Maribeth Gandy}.} \bibinfo{year}{2005}\natexlab{}.
\newblock \showarticletitle{Wizard of Oz support throughout an iterative design process}.
\newblock \bibinfo{journal}{\emph{IEEE Pervasive Computing}} \bibinfo{volume}{4}, \bibinfo{number}{4} (\bibinfo{year}{2005}), \bibinfo{pages}{18--26}.
\newblock


\bibitem[Eckhardt et~al\mbox{.}(2023)]%
        {Eckhardt2023}
\bibfield{author}{\bibinfo{person}{Sven Eckhardt}, \bibinfo{person}{Merlin Knaeble}, \bibinfo{person}{Andreas Bucher}, \bibinfo{person}{Dario Staehelin}, \bibinfo{person}{Mateusz Dolata}, \bibinfo{person}{Doris Agotai}, {and} \bibinfo{person}{Gerhard Schwabe}.} \bibinfo{year}{2023}\natexlab{}.
\newblock \bibinfo{booktitle}{\emph{“Garbage In, Garbage Out”: Mitigating Human Biases in Data Entry by Means of Artificial Intelligence}}.
\newblock \bibinfo{publisher}{Springer Nature Switzerland}, \bibinfo{pages}{27–48}.
\newblock
\showISBNx{9783031422867}
\showISSN{1611-3349}
\urldef\tempurl%
\url{https://doi.org/10.1007/978-3-031-42286-7_2}
\showDOI{\tempurl}


\bibitem[Finn and Louviere(1992)]%
        {Finn1992}
\bibfield{author}{\bibinfo{person}{Adam Finn} {and} \bibinfo{person}{Jordan~J. Louviere}.} \bibinfo{year}{1992}\natexlab{}.
\newblock \showarticletitle{Determining the Appropriate Response to Evidence of Public Concern: The Case of Food Safety}.
\newblock \bibinfo{journal}{\emph{Journal of Public Policy {\&} Marketing}} \bibinfo{volume}{11}, \bibinfo{number}{2} (\bibinfo{date}{Sept.} \bibinfo{year}{1992}), \bibinfo{pages}{12--25}.
\newblock
\urldef\tempurl%
\url{https://doi.org/10.1177/074391569201100202}
\showDOI{\tempurl}


\bibitem[Fraser and Gilbert(1991)]%
        {fraser1991simulating}
\bibfield{author}{\bibinfo{person}{Norman~M Fraser} {and} \bibinfo{person}{G~Nigel Gilbert}.} \bibinfo{year}{1991}\natexlab{}.
\newblock \showarticletitle{Simulating speech systems}.
\newblock \bibinfo{journal}{\emph{Computer Speech \& Language}} \bibinfo{volume}{5}, \bibinfo{number}{1} (\bibinfo{year}{1991}), \bibinfo{pages}{81--99}.
\newblock


\bibitem[Galitsky(2023)]%
        {galitsky2023truth}
\bibfield{author}{\bibinfo{person}{Boris~A Galitsky}.} \bibinfo{year}{2023}\natexlab{}.
\newblock \bibinfo{title}{Truth-O-Meter: Collaborating with LLM in Fighting its Hallucinations}.
\newblock
\newblock


\bibitem[Gao et~al\mbox{.}(2023)]%
        {gao2023rarr}
\bibfield{author}{\bibinfo{person}{Luyu Gao}, \bibinfo{person}{Zhuyun Dai}, \bibinfo{person}{Panupong Pasupat}, \bibinfo{person}{Anthony Chen}, \bibinfo{person}{Arun~Tejasvi Chaganty}, \bibinfo{person}{Yicheng Fan}, \bibinfo{person}{Vincent~Y. Zhao}, \bibinfo{person}{Ni Lao}, \bibinfo{person}{Hongrae Lee}, \bibinfo{person}{Da-Cheng Juan}, {and} \bibinfo{person}{Kelvin Guu}.} \bibinfo{year}{2023}\natexlab{}.
\newblock \bibinfo{title}{RARR: Researching and Revising What Language Models Say, Using Language Models}.
\newblock
\newblock


\bibitem[Google(2021)]%
        {lamda2021}
\bibfield{author}{\bibinfo{person}{Google}.} \bibinfo{year}{2021}\natexlab{}.
\newblock \bibinfo{title}{{LaMDA}: our breakthrough conversation technology}.
\newblock
\newblock
\urldef\tempurl%
\url{https://blog.google/technology/ai/lamda/}
\showURL{%
\tempurl}


\bibitem[Guerreiro et~al\mbox{.}(2023)]%
        {guerreiro2023hallucinations}
\bibfield{author}{\bibinfo{person}{Nuno~M. Guerreiro}, \bibinfo{person}{Duarte Alves}, \bibinfo{person}{Jonas Waldendorf}, \bibinfo{person}{Barry Haddow}, \bibinfo{person}{Alexandra Birch}, \bibinfo{person}{Pierre Colombo}, {and} \bibinfo{person}{André F.~T. Martins}.} \bibinfo{year}{2023}\natexlab{}.
\newblock \bibinfo{title}{Hallucinations in Large Multilingual Translation Models}.
\newblock
\newblock
\showeprint[arxiv]{2303.16104}~[cs.CL]


\bibitem[Holzinger(2018)]%
        {holzinger2018machine}
\bibfield{author}{\bibinfo{person}{Andreas Holzinger}.} \bibinfo{year}{2018}\natexlab{}.
\newblock \showarticletitle{From machine learning to explainable AI}. In \bibinfo{booktitle}{\emph{2018 World Symposium on Digital Intelligence for Systems and Machines (DISA)}}. IEEE, \bibinfo{pages}{55--66}.
\newblock


\bibitem[Ji et~al\mbox{.}(2023)]%
        {ji2023survey}
\bibfield{author}{\bibinfo{person}{Ziwei Ji}, \bibinfo{person}{Nayeon Lee}, \bibinfo{person}{Rita Frieske}, \bibinfo{person}{Tiezheng Yu}, \bibinfo{person}{Dan Su}, \bibinfo{person}{Yan Xu}, \bibinfo{person}{Etsuko Ishii}, \bibinfo{person}{Ye~Jin Bang}, \bibinfo{person}{Andrea Madotto}, {and} \bibinfo{person}{Pascale Fung}.} \bibinfo{year}{2023}\natexlab{}.
\newblock \showarticletitle{Survey of hallucination in natural language generation}.
\newblock \bibinfo{journal}{\emph{Comput. Surveys}} \bibinfo{volume}{55}, \bibinfo{number}{12} (\bibinfo{year}{2023}), \bibinfo{pages}{1--38}.
\newblock


\bibitem[Kahne and Bowyer(2018)]%
        {kahne2018political}
\bibfield{author}{\bibinfo{person}{Joseph Kahne} {and} \bibinfo{person}{Benjamin Bowyer}.} \bibinfo{year}{2018}\natexlab{}.
\newblock \showarticletitle{The political significance of social media activity and social networks}.
\newblock \bibinfo{journal}{\emph{Political Communication}} \bibinfo{volume}{35}, \bibinfo{number}{3} (\bibinfo{year}{2018}), \bibinfo{pages}{470--493}.
\newblock


\bibitem[Lai et~al\mbox{.}(2021)]%
        {lai2021towards}
\bibfield{author}{\bibinfo{person}{Vivian Lai}, \bibinfo{person}{Chacha Chen}, \bibinfo{person}{Q~Vera Liao}, \bibinfo{person}{Alison Smith-Renner}, {and} \bibinfo{person}{Chenhao Tan}.} \bibinfo{year}{2021}\natexlab{}.
\newblock \bibinfo{title}{Towards a science of human-ai decision making: a survey of empirical studies}.
\newblock
\newblock


\bibitem[Lee and See(2004)]%
        {lee2004trust}
\bibfield{author}{\bibinfo{person}{John~D Lee} {and} \bibinfo{person}{Katrina~A See}.} \bibinfo{year}{2004}\natexlab{}.
\newblock \showarticletitle{Trust in automation: Designing for appropriate reliance}.
\newblock \bibinfo{journal}{\emph{Human Factors}} \bibinfo{volume}{46}, \bibinfo{number}{1} (\bibinfo{year}{2004}), \bibinfo{pages}{50--80}.
\newblock


\bibitem[Leiser et~al\mbox{.}(2023)]%
        {leiser_chatgpt_2023}
\bibfield{author}{\bibinfo{person}{Florian Leiser}, \bibinfo{person}{Sven Eckhardt}, \bibinfo{person}{Merlin Knaeble}, \bibinfo{person}{Alexander Maedche}, \bibinfo{person}{Gerhard Schwabe}, {and} \bibinfo{person}{Ali Sunyaev}.} \bibinfo{year}{2023}\natexlab{}.
\newblock \showarticletitle{From {ChatGPT} to {FactGPT}: {A} {Participatory} {Design} {Study} to {Mitigate} the {Effects} of {Large} {Language} {Model} {Hallucinations} on {Users}}. In \bibinfo{booktitle}{\emph{Proceedings of {Mensch} und {Computer} 2023 ({MuC} '23)}}. \bibinfo{publisher}{ACM}, \bibinfo{address}{Rapperswil, Switzerland}, \bibinfo{pages}{81--90}.
\newblock
\showISBNx{979-8-4007-0771-1/23/09}


\bibitem[Manakul et~al\mbox{.}(2023)]%
        {manakul2023selfcheckgpt}
\bibfield{author}{\bibinfo{person}{Potsawee Manakul}, \bibinfo{person}{Adian Liusie}, {and} \bibinfo{person}{Mark~JF Gales}.} \bibinfo{year}{2023}\natexlab{}.
\newblock \showarticletitle{Selfcheckgpt: Zero-resource black-box hallucination detection for generative large language models}.
\newblock \bibinfo{journal}{\emph{arXiv preprint arXiv:2303.08896}} (\bibinfo{year}{2023}).
\newblock


\bibitem[Maynez et~al\mbox{.}(2020)]%
        {maynez2020faithfulness}
\bibfield{author}{\bibinfo{person}{Joshua Maynez}, \bibinfo{person}{Shashi Narayan}, \bibinfo{person}{Bernd Bohnet}, {and} \bibinfo{person}{Ryan McDonald}.} \bibinfo{year}{2020}\natexlab{}.
\newblock \showarticletitle{On faithfulness and factuality in abstractive summarization}.
\newblock \bibinfo{journal}{\emph{arXiv preprint arXiv:2005.00661}} (\bibinfo{year}{2020}).
\newblock


\bibitem[Myers and Newman(2007)]%
        {myers2007qualitative}
\bibfield{author}{\bibinfo{person}{Michael~D Myers} {and} \bibinfo{person}{Michael Newman}.} \bibinfo{year}{2007}\natexlab{}.
\newblock \showarticletitle{The qualitative interview in IS research: Examining the craft}.
\newblock \bibinfo{journal}{\emph{Information and Organization}} \bibinfo{volume}{17}, \bibinfo{number}{1} (\bibinfo{year}{2007}), \bibinfo{pages}{2--26}.
\newblock


\bibitem[Narayan et~al\mbox{.}(2023)]%
        {narayan_elon_2023}
\bibfield{author}{\bibinfo{person}{Jyoti Narayan}, \bibinfo{person}{Krystal Hu}, \bibinfo{person}{Martin Coulter}, {and} \bibinfo{person}{Supantha Mukherjee}.} \bibinfo{year}{2023}\natexlab{}.
\newblock \bibinfo{title}{Elon {Musk} and others urge ai pause, citing 'risks to society'}.
\newblock
\newblock
\urldef\tempurl%
\url{https://www.reuters.com/technology/musk-experts-urge-pause-training-ai-systems-that-can-outperform-gpt-4-2023-03-29/}
\showURL{%
\tempurl}
\newblock
\shownote{Publication Title: Reuters}.


\bibitem[OpenAI(2023a)]%
        {openai2023gpt4}
\bibfield{author}{\bibinfo{person}{OpenAI}.} \bibinfo{year}{2023}\natexlab{a}.
\newblock \bibinfo{title}{GPT-4 Technical Report}.
\newblock
\newblock
\showeprint[arxiv]{2303.08774}~[cs.CL]


\bibitem[OpenAI(2023b)]%
        {openai2023}
\bibfield{author}{\bibinfo{person}{OpenAI}.} \bibinfo{year}{2023}\natexlab{b}.
\newblock \bibinfo{title}{Introducing {ChatGPT}}.
\newblock
\newblock
\urldef\tempurl%
\url{https://openai.com/blog/chatgpt}
\showURL{%
\tempurl}


\bibitem[Ouyang et~al\mbox{.}(2022)]%
        {NEURIPS2022_Ouyang}
\bibfield{author}{\bibinfo{person}{Long Ouyang}, \bibinfo{person}{Jeffrey Wu}, \bibinfo{person}{Xu Jiang}, \bibinfo{person}{Diogo Almeida}, \bibinfo{person}{Carroll Wainwright}, \bibinfo{person}{Pamela Mishkin}, \bibinfo{person}{Chong Zhang}, \bibinfo{person}{Sandhini Agarwal}, \bibinfo{person}{Katarina Slama}, \bibinfo{person}{Alex Ray}, \bibinfo{person}{John Schulman}, \bibinfo{person}{Jacob Hilton}, \bibinfo{person}{Fraser Kelton}, \bibinfo{person}{Luke Miller}, \bibinfo{person}{Maddie Simens}, \bibinfo{person}{Amanda Askell}, \bibinfo{person}{Peter Welinder}, \bibinfo{person}{Paul~F Christiano}, \bibinfo{person}{Jan Leike}, {and} \bibinfo{person}{Ryan Lowe}.} \bibinfo{year}{2022}\natexlab{}.
\newblock \showarticletitle{Training language models to follow instructions with human feedback}. In \bibinfo{booktitle}{\emph{Advances in Neural Information Processing Systems}}, \bibfield{editor}{\bibinfo{person}{S.~Koyejo}, \bibinfo{person}{S.~Mohamed}, \bibinfo{person}{A.~Agarwal}, \bibinfo{person}{D.~Belgrave}, \bibinfo{person}{K.~Cho}, {and} \bibinfo{person}{A.~Oh}} (Eds.), Vol.~\bibinfo{volume}{35}. \bibinfo{publisher}{Curran Associates, Inc.}, \bibinfo{pages}{27730--27744}.
\newblock


\bibitem[Parasuraman and Riley(1997)]%
        {parasuraman1997humans}
\bibfield{author}{\bibinfo{person}{Raja Parasuraman} {and} \bibinfo{person}{Victor Riley}.} \bibinfo{year}{1997}\natexlab{}.
\newblock \showarticletitle{Humans and automation: Use, misuse, disuse, abuse}.
\newblock \bibinfo{journal}{\emph{Human factors}} \bibinfo{volume}{39}, \bibinfo{number}{2} (\bibinfo{year}{1997}), \bibinfo{pages}{230--253}.
\newblock


\bibitem[Passi and Vorvoreanu(2022)]%
        {passi2022overreliance}
\bibfield{author}{\bibinfo{person}{Samir Passi} {and} \bibinfo{person}{Mihaela Vorvoreanu}.} \bibinfo{year}{2022}\natexlab{}.
\newblock \showarticletitle{Overreliance on AI Literature Review}.
\newblock \bibinfo{journal}{\emph{Microsoft Research}} (\bibinfo{year}{2022}).
\newblock


\bibitem[Peng et~al\mbox{.}(2023)]%
        {peng2023check}
\bibfield{author}{\bibinfo{person}{Baolin Peng}, \bibinfo{person}{Michel Galley}, \bibinfo{person}{Pengcheng He}, \bibinfo{person}{Hao Cheng}, \bibinfo{person}{Yujia Xie}, \bibinfo{person}{Yu Hu}, \bibinfo{person}{Qiuyuan Huang}, \bibinfo{person}{Lars Liden}, \bibinfo{person}{Zhou Yu}, \bibinfo{person}{Weizhu Chen}, {et~al\mbox{.}}} \bibinfo{year}{2023}\natexlab{}.
\newblock \showarticletitle{Check your facts and try again: Improving large language models with external knowledge and automated feedback}.
\newblock \bibinfo{journal}{\emph{arXiv preprint arXiv:2302.12813}} (\bibinfo{year}{2023}).
\newblock


\bibitem[Rajpurkar et~al\mbox{.}(2018)]%
        {squad_2018_rajpurkar}
\bibfield{author}{\bibinfo{person}{Pranav Rajpurkar}, \bibinfo{person}{Robin Jia}, {and} \bibinfo{person}{Percy Liang}.} \bibinfo{year}{2018}\natexlab{}.
\newblock \showarticletitle{Know What You Don't Know: Unanswerable Questions for SQuAD}.
\newblock \bibinfo{journal}{\emph{CoRR}}  \bibinfo{volume}{abs/1806.03822} (\bibinfo{year}{2018}).
\newblock
\showeprint[arXiv]{1806.03822}
\urldef\tempurl%
\url{http://arxiv.org/abs/1806.03822}
\showURL{%
\tempurl}


\bibitem[Raunak et~al\mbox{.}(2021)]%
        {raunak2021curious}
\bibfield{author}{\bibinfo{person}{Vikas Raunak}, \bibinfo{person}{Arul Menezes}, {and} \bibinfo{person}{Marcin Junczys-Dowmunt}.} \bibinfo{year}{2021}\natexlab{}.
\newblock \bibinfo{title}{The Curious Case of Hallucinations in Neural Machine Translation}.
\newblock
\newblock
\showeprint[arxiv]{2104.06683}~[cs.CL]


\bibitem[Riek(2012)]%
        {riek2012wizard}
\bibfield{author}{\bibinfo{person}{Laurel~D Riek}.} \bibinfo{year}{2012}\natexlab{}.
\newblock \showarticletitle{Wizard of Oz studies in HRI: A Systematic Review and New Reporting Guidelines}.
\newblock \bibinfo{journal}{\emph{Journal of Human-Robot Interaction}} \bibinfo{volume}{1}, \bibinfo{number}{1} (\bibinfo{year}{2012}), \bibinfo{pages}{119--136}.
\newblock


\bibitem[Schade(2021)]%
        {openaiWhatDoes}
\bibfield{author}{\bibinfo{person}{Michael Schade}.} \bibinfo{year}{2021}\natexlab{}.
\newblock \bibinfo{title}{{W}hat {D}oes the {O}fficial {C}hat{G}{P}{T} i{O}{S} {A}pp {I}con {L}ook {L}ike? | {O}pen{A}{I} {H}elp {C}enter --- help.openai.com}.
\newblock \bibinfo{howpublished}{\url{https://help.openai.com/en/articles/7905742-what-does-the-official-chatgpt-ios-app-icon-look-like}}.
\newblock
\newblock
\shownote{[Accessed 25-08-2023]}.


\bibitem[Schemmer et~al\mbox{.}(2023)]%
        {schemmer2023appropriate}
\bibfield{author}{\bibinfo{person}{Max Schemmer}, \bibinfo{person}{Niklas Kuehl}, \bibinfo{person}{Carina Benz}, \bibinfo{person}{Andrea Bartos}, {and} \bibinfo{person}{Gerhard Satzger}.} \bibinfo{year}{2023}\natexlab{}.
\newblock \showarticletitle{Appropriate reliance on AI advice: Conceptualization and the effect of explanations}. In \bibinfo{booktitle}{\emph{Proceedings of the 28th International Conference on Intelligent User Interfaces}}. \bibinfo{pages}{410--422}.
\newblock


\bibitem[Shieh(2023)]%
        {openaiBestPractices}
\bibfield{author}{\bibinfo{person}{Jessica Shieh}.} \bibinfo{year}{2023}\natexlab{}.
\newblock \bibinfo{title}{{B}est practices for prompt engineering with {O}pen{A}{I} {A}{P}{I} | {O}pen{A}{I} {H}elp {C}enter --- help.openai.com}.
\newblock \bibinfo{howpublished}{\url{https://help.openai.com/en/articles/6654000-best-practices-for-prompt-engineering-with-openai-api}}.
\newblock
\newblock
\shownote{[Accessed 29-08-2023]}.


\bibitem[Shneiderman(2020)]%
        {shneiderman2020human}
\bibfield{author}{\bibinfo{person}{Ben Shneiderman}.} \bibinfo{year}{2020}\natexlab{}.
\newblock \showarticletitle{Human-centered artificial intelligence: Reliable, safe \& trustworthy}.
\newblock \bibinfo{journal}{\emph{International Journal of Human-Computer Interaction}} \bibinfo{volume}{36}, \bibinfo{number}{6} (\bibinfo{year}{2020}), \bibinfo{pages}{495--504}.
\newblock


\bibitem[Thiebes et~al\mbox{.}(2020)]%
        {Thiebes2020}
\bibfield{author}{\bibinfo{person}{Scott Thiebes}, \bibinfo{person}{Sebastian Lins}, {and} \bibinfo{person}{Ali Sunyaev}.} \bibinfo{year}{2020}\natexlab{}.
\newblock \showarticletitle{Trustworthy artificial intelligence}.
\newblock \bibinfo{journal}{\emph{Electronic Markets}} \bibinfo{volume}{31}, \bibinfo{number}{2} (\bibinfo{date}{Oct.} \bibinfo{year}{2020}), \bibinfo{pages}{447–464}.
\newblock
\showISSN{1422-8890}
\urldef\tempurl%
\url{https://doi.org/10.1007/s12525-020-00441-4}
\showDOI{\tempurl}


\bibitem[Tomitza et~al\mbox{.}(ming)]%
        {tomitza2023minimum}
\bibfield{author}{\bibinfo{person}{Christoph Tomitza}, \bibinfo{person}{Myriam Schaschek}, \bibinfo{person}{Lisa Straub}, {and} \bibinfo{person}{Axel Winkelmann}.} \bibinfo{year}{forthcoming}\natexlab{}.
\newblock \showarticletitle{What is the Minimum to Trust AI?—A Requirement Analysis for (Generative) AI-based Texts}. In \bibinfo{booktitle}{\emph{Internationale Tagung Wirtschaftsinformatik 2023}}.
\newblock


\bibitem[Vincent(2023)]%
        {vincent_google_2023}
\bibfield{author}{\bibinfo{person}{James Vincent}.} \bibinfo{year}{2023}\natexlab{}.
\newblock \bibinfo{title}{Google and {Microsoft}’s chatbots are already citing one another in a misinformation shitshow}.
\newblock
\newblock
\urldef\tempurl%
\url{https://www.theverge.com/2023/3/22/23651564/google-microsoft-bard-bing-chatbots-misinformation}
\showURL{%
\tempurl}


\bibitem[von Rueden et~al\mbox{.}(2023)]%
        {vonrueden2023informed}
\bibfield{author}{\bibinfo{person}{Laura von Rueden}, \bibinfo{person}{Sebastian Mayer}, \bibinfo{person}{Katharina Beckh}, \bibinfo{person}{Bogdan Georgiev}, \bibinfo{person}{Sven Giesselbach}, \bibinfo{person}{Raoul Heese}, \bibinfo{person}{Birgit Kirsch}, \bibinfo{person}{Julius Pfrommer}, \bibinfo{person}{Annika Pick}, \bibinfo{person}{Rajkumar Ramamurthy}, \bibinfo{person}{Michal Walczak}, \bibinfo{person}{Jochen Garcke}, \bibinfo{person}{Christian Bauckhage}, {and} \bibinfo{person}{Jannis Schuecker}.} \bibinfo{year}{2023}\natexlab{}.
\newblock \showarticletitle{Informed Machine Learning – A Taxonomy and Survey of Integrating Prior Knowledge into Learning Systems}.
\newblock \bibinfo{journal}{\emph{IEEE Transactions on Knowledge and Data Engineering}} \bibinfo{volume}{35}, \bibinfo{number}{1} (\bibinfo{year}{2023}), \bibinfo{pages}{614--633}.
\newblock
\urldef\tempurl%
\url{https://doi.org/10.1109/TKDE.2021.3079836}
\showDOI{\tempurl}


\bibitem[White et~al\mbox{.}(2023)]%
        {white2023prompt}
\bibfield{author}{\bibinfo{person}{Jules White}, \bibinfo{person}{Quchen Fu}, \bibinfo{person}{Sam Hays}, \bibinfo{person}{Michael Sandborn}, \bibinfo{person}{Carlos Olea}, \bibinfo{person}{Henry Gilbert}, \bibinfo{person}{Ashraf Elnashar}, \bibinfo{person}{Jesse Spencer-Smith}, {and} \bibinfo{person}{Douglas~C. Schmidt}.} \bibinfo{year}{2023}\natexlab{}.
\newblock \bibinfo{title}{A Prompt Pattern Catalog to Enhance Prompt Engineering with ChatGPT}.
\newblock
\newblock
\showeprint[arxiv]{2302.11382}~[cs.SE]


\bibitem[Yang et~al\mbox{.}(2023)]%
        {yang_2023_LLMsinPractice}
\bibfield{author}{\bibinfo{person}{Jingfeng Yang}, \bibinfo{person}{Hongye Jin}, \bibinfo{person}{Ruixiang Tang}, \bibinfo{person}{Xiaotian Han}, \bibinfo{person}{Qizhang Feng}, \bibinfo{person}{Haoming Jiang}, \bibinfo{person}{Bing Yin}, {and} \bibinfo{person}{Xia Hu}.} \bibinfo{year}{2023}\natexlab{}.
\newblock \bibinfo{title}{Harnessing the Power of LLMs in Practice: A Survey on ChatGPT and Beyond}.
\newblock
\newblock
\urldef\tempurl%
\url{https://doi.org/10.48550/ARXIV.2304.13712}
\showDOI{\tempurl}


\bibitem[Zamfirescu-Pereira et~al\mbox{.}(2023)]%
        {zamfirescu2023johnny}
\bibfield{author}{\bibinfo{person}{JD Zamfirescu-Pereira}, \bibinfo{person}{Richmond~Y Wong}, \bibinfo{person}{Bjoern Hartmann}, {and} \bibinfo{person}{Qian Yang}.} \bibinfo{year}{2023}\natexlab{}.
\newblock \showarticletitle{Why Johnny can’t prompt: how non-AI experts try (and fail) to design LLM prompts}. In \bibinfo{booktitle}{\emph{Proceedings of the 2023 CHI Conference on Human Factors in Computing Systems}}. \bibinfo{pages}{1--21}.
\newblock


\bibitem[Zhang et~al\mbox{.}(2023)]%
        {zhang2023siren}
\bibfield{author}{\bibinfo{person}{Yue Zhang}, \bibinfo{person}{Yafu Li}, \bibinfo{person}{Leyang Cui}, \bibinfo{person}{Deng Cai}, \bibinfo{person}{Lemao Liu}, \bibinfo{person}{Tingchen Fu}, \bibinfo{person}{Xinting Huang}, \bibinfo{person}{Enbo Zhao}, \bibinfo{person}{Yu Zhang}, \bibinfo{person}{Yulong Chen}, {et~al\mbox{.}}} \bibinfo{year}{2023}\natexlab{}.
\newblock \showarticletitle{Siren's Song in the AI Ocean: A Survey on Hallucination in Large Language Models}.
\newblock \bibinfo{journal}{\emph{arXiv preprint arXiv:2309.01219}} (\bibinfo{year}{2023}).
\newblock


\bibitem[Zhao et~al\mbox{.}(2020)]%
        {zhao2020reducing}
\bibfield{author}{\bibinfo{person}{Zheng Zhao}, \bibinfo{person}{Shay~B. Cohen}, {and} \bibinfo{person}{Bonnie Webber}.} \bibinfo{year}{2020}\natexlab{}.
\newblock \bibinfo{title}{Reducing Quantity Hallucinations in Abstractive Summarization}.
\newblock
\newblock
\showeprint[arxiv]{2009.13312}~[cs.CL]


\bibitem[Zhou et~al\mbox{.}(2023)]%
        {zhou2023analyzing}
\bibfield{author}{\bibinfo{person}{Yiyang Zhou}, \bibinfo{person}{Chenhang Cui}, \bibinfo{person}{Jaehong Yoon}, \bibinfo{person}{Linjun Zhang}, \bibinfo{person}{Zhun Deng}, \bibinfo{person}{Chelsea Finn}, \bibinfo{person}{Mohit Bansal}, {and} \bibinfo{person}{Huaxiu Yao}.} \bibinfo{year}{2023}\natexlab{}.
\newblock \bibinfo{title}{Analyzing and Mitigating Object Hallucination in Large Vision-Language Models}.
\newblock
\newblock
\showeprint[arxiv]{2310.00754}~[cs.LG]


\end{thebibliography}
\end{document}